\pdfoutput=1
\documentclass{article}%
\usepackage{amsmath}
\usepackage{amsfonts}
\usepackage{amssymb}
\usepackage{graphicx}%
\newtheorem{theorem}{Theorem}

\newtheorem{remark}[theorem]{Remark}

\newcommand{\bpartial}{\mathop{\partial\kern -4pt\raisebox{.8pt}{$|$}}}
\newcommand{\sbpartial}{\tiny\mathop{\partial\kern -4pt\raisebox{.8pt}{$|$}}}
\newcommand{\bra}{\mathopen{[\kern-1.6pt[}}
\newcommand{\ket}{\mathclose{]\kern-1.5pt]}}
\newcommand{\bbra}{\mathopen{[\kern-2.2pt[\kern-2.3pt[}}
\newcommand{\bket}{\mathclose{]\kern-2.1pt]\kern-2.3pt]}}
\newcommand{\sla}{\mbox{\bfseries\slshape a}}
\newcommand{\sle}{\mbox{\bfseries\slshape e}}
\newcommand{\slg}{\mbox{\bfseries\slshape g}}
\newcommand{\sslg}{\mbox{\tiny \bfseries\slshape g}}

\makeindex
\begin{document}

\title{Some Thoughts on Geometries and on the Nature of the Gravitational Field}
\author{{\footnotesize Eduardo A. Notte-Cuello}$^{(1)}${\footnotesize , Rold\~{a}o da
Rocha }$^{(2)}${\footnotesize and Waldyr A. Rodrigues Jr.}$^{(3)}$\\$^{(1)}${\small Departamento de} {\small Matem\'{a}ticas,Universidad de La
Serena,}\\{\small Av. Cisternas 1200, La Serena-Chile}\\$^{(2)}${\footnotesize Centro de Matem\'{a}tica, Computa\c{c}\~{a}o e
Cogni\c{c}\~{a}o}\\{\footnotesize Universidade Federal do ABC, 09210-170, Santo Andr\'{e}, SP,
Brazil}\\$^{(3)}\,\hspace{-0.1cm}${\footnotesize Institute of Mathematics, Statistics
and Scientific Computation}\\{\footnotesize IMECC-UNICAMP CP 6065}\\{\footnotesize 13083-859 Campinas, SP, Brazil}\\{\small e-mail:} {\small enotte@userena.cl;
{\footnotesize roldao.rocha@ufabc.edu.br;} walrod@ime.unicamp.br}}
\maketitle

\begin{abstract}
In this paper we show how a gravitational field generated by a given
energy-momentum distribution (for all realistic cases) can be represented by
distinct geometrical structures (Lorentzian, teleparallel and non null
nonmetricity spacetimes) or that we even can dispense all those geometrical
structures and simply represent the gravitational field as a field in the
Faraday's sense living in Minkowski spacetime. The explicit Lagrangian density
for this theory is given and the field equations (which are Maxwell's like
equations) are shown to be equivalent to Einstein's equations. Some examples
are worked in detail in order to convince the reader that the geometrical
structure of a manifold (modulus some topological constraints) is conventional
as already emphasized by Poincar\'{e} long ago, and thus the realization that
there are disctints geometrical representations (and a physical model related
to a deformation of the continuum supporting Minkowski spacetime) for any
realistic gravitational field strongly suggests that we must investigate the
origin of its physical nature. We hope that this paper will convince readers
that this is indeed the case.

\end{abstract}

\section{Introduction}

Physics students learn General Relativity (GR) as the modern theory of
gravitation. In that theory each gravitational field generated by a given
energy-momentum tensor is represented by a Lorentzian spacetime, i.e., a
structure $(M,D,%
\slg
,\tau_{%
\sslg
},\uparrow)$ where $\ M$ is a non compact (locally compact) $4$-dimensional
Hausdorff manifold, $%
\slg
$ is a Lorentzian metric on $M$ and $D$ is its Levi-Civita connection.
Moreover $M$ is supposed oriented by the volume form\ $\tau_{%
\sslg
\ }$ and the symbol $\uparrow$ means that the spacetime is time
orientable\footnote{For details, please consult, e.g., \cite{rodcap2007,sawu}%
.}. From the geometrical objects in the structure $(M,D,%
\slg
,\tau_{%
\sslg
},\uparrow)$ we can calculate the \textit{Riemann curvature tensor}
$\mathbf{R}$ of $D$ and a nontrivial GR model is one in which $\mathbf{R}%
\neq0$. In that way textbooks often say that in GR \textit{spacetime is
curved}. Unfortunately many people mislead the curvature of a connection $D$
on $M$ with the fact that $M$ can eventually be a bent surface in an (pseudo)
Euclidean space with a sufficient number of dimensions\footnote{Any manifold
$M,\dim M=n$ according to the Whitney theorem can be realized as a submanifold
of $\mathbb{R}^{m}$, with $m=2n$. However, if $M$ carries additional structure
the number $m$ in general must be greater than $2n$. Indeed, it has been shown
by Eddington \cite{eddington} that if dim $M=4$ and if $M$ carries a
Lorentzian metric $%
\slg
$, which moreover satisfies Einstein's equations, then $M$ can be locally
embedded in a (pseudo)euclidean space $\mathbb{R}^{1,9}$ Also, isometric
embeddings of general Lorentzian spacetimes would require a lot of extra
dimensions \cite{clarke}. Indeed, a compact Lorentzian manifold can be
embedded isometrically in $\mathbb{R}^{2,46}$ and a non-compact one can be
embedded isometrically in $\mathbb{R}^{2,87}$!}. This confusion leads to all
sort of wishful thinking because many forget that GR does not fix the
topology\footnote{In particular the topology of the universe that we live in
is not known, as yet \cite{weeks}.} of $M$ that often must be put
\textquotedblleft by hand\textquotedblright\ when solving a problem,\ and thus
think that they can bend spacetime if they have an appropriate kind of some
exotic matter. Worse, the insistence in supposing that the gravitational field
is \textit{geometry} lead the majority of physicists to relegate the search
for the real physical nature of the gravitational field as not important at
all\footnote{See a nice discussion of this issue in \cite{laughlin}.}.Instead,
students are advertised that GR is considered by may physicists as the most
beautiful physical theory \cite{landau}. However textbooks with a few
exceptions (see, e.g., the excellent book by Sachs and Wu \cite{sawu}) forget
to say to their readers that in GR there are no genuine conservation laws of
energy-momentum and angular momentum unless spacetime has some
\textit{additional} structure which is not present in a general Lorentzian
spacetime \cite{notterod1}. Only a few people tried to develop consistently
theories where the gravitational field (at least from the classical point of
view) is a field in the Faraday's sense living in Minkowski spacetime (see below).

In this paper we want to recall two important results that hopefully will lead
people to realize that eventually it is time to disclose the real nature of
the gravitational field\footnote{Of course, the authors know that one of the
claims of string theory is that it nicely describes gravitation. In that
theory General Relativity is only an approximated theory valid for distances
much greater than the Planck length.}. The first result is that the
representation of gravitational fields by Lorentzian spacetimes is eventually
no more than an consequence of to the differential geometry knowledge of
Einstein and Grossmann when they where struggling to find a consistent way to
describe the gravitational field\footnote{See some detais below.}. Indeed,
there are some geometrical structures different from $(M,D,%
\slg
,\tau_{%
\sslg
},\uparrow)$ that can equivalently represent such a field. The second result
is that the gravitational field (in all known situations) can also be nicely
represented as a field in the Faraday's sense \cite{notterod} living in a
fixed background spacetime\footnote{The preferred one is, of course, Minkowski
spacetime, the simple choice. But, the true background spacetime may be
eventually a more complicated one, since that manifold must represent the
global topological structure of the universe, something that is not known at
the time of this writing \cite{weeks}.}. Concerning the alternative
geometrical models, the particular cases where the connection is teleparallel
(i.e., it is metrical compatible, has \textit{null} Riemann curvature tensor
and \textit{non null }torsion tensor) and the one where the connection is
\textit{not} metrical compatible (i.e., its nonmetricity tensor $\mathbf{A}%
^{\eta}\neq0$) will be addressed below. However to understand how those
alternative geometrical models (and the physical model) can be constructed
\ and why the Lorentzian spacetime model was Einstein's first choice, it is
eventually worth to recall some historical facts concerning attempts by
Einstein (and others) to build a geometrical unified theory of the
gravitational and electromagnetic fields.

We start with the word torsion. Although such a word seems to have been
introduced by Cartan \cite{cartan} in 1922 the fact is that the concept behind
the name already appeared in a Ricci's paper \cite{ricci1} from 1895 and was
also used in \cite{ricci2}! In those papers Ricci introduced what is now
called the Cartan's\textit{ moving frames} and the \textit{teleparallel
geometry\footnote{Also known as Weintzb\"{o}ck geometry \cite{wein}.}}.

Moreover in 1901 Ricci and Levi-Civita\footnote{A Ricci's student at that
time.} published a joy \cite{riccicivita}, which has become the \textit{bible}
of tensor calculus and which has been extensively studied by Einstein and
Grossmann in their search for the theory of the gravitational
field\footnote{An english translation of the joy with very useful comments has
been done by the mathematical physicist Robert Hermann \cite{hermann} in 1975
and that text (and many others books by Hermann) can be downloaded from
http://books.google.com.br/books?q=robert+hermann}. However Einstein and
Grossmann seems to have studied only the first part of reference 4 and so
missed the \textquotedblleft Cartan's moving method\textquotedblright\ and the
concept of torsion. It seems also that only after 1922 Einstein become
interested in the second chapter of the joy, titled \textit{La
G\'{e}om\'{e}trie Intrinseque Comme Instrument de Calcul} and discovered
torsion and the teleparallel geometry\footnote{Some interesting historical
details may be found in \cite{goenner,schu}}. As it is well known he tried to
identify a certain contraction of the torsion tensor of a teleparallel
geometry with the electromagnetic potential, but after sometime he discovered
that the idea did not work. Einstein's first papers on the
subject\footnote{For a complete list of Einstein's papers on the subject see
\cite{goenner}.} are \cite{einstein2,einstein3,einstein4}. Also in a paper
which Einstein wrote in 1925 \cite{einstein1} the torsion tensor concept
already appeared, since he considered as one of his field variables the
antisymmetric part of a \textit{non symmetric} connection. All those papers by
Einstein have been translated into English by Unzicker and Case
\cite{unzicker} and can be downloaded from the arXiv. We also can learn in
\cite{debewer,goenner} that Cartan tried to explain the teleparallel geometry
to Einstein when he visited Paris in 1922 using the example of what we call
the \textit{Nunes connection }(\textit{or navigator connection}) on the
punctured sphere. Since this example illustrates in a crystal clear way the
fact that one must not confound the Riemann curvature of a given connection
defined on a manifold $M$ with the fact that $M$ may be viewed as a
\textit{bent} hypersurface embedded in an Euclidean space (with appropriate
number of dimensions) it will be presented in Appendix A. A comparison of the
parallel transport according to the Nunes connection and according to the
usual Levi-Civita connection is done\footnote{The material of Appendix A
follows the presentation in Section 4.7.7 of \cite{rodcap2007}.}, and it is
shown that the Nunes connection the Riemann curvature of the punctured sphere
is \textit{null}. In this sense the geometry of the punctured sphere is
conventional as emphasized by Poincar\'{e} \cite{poincare} long ago.

As we already said the main objective of the present paper is to clarify the
fact that there are different ways of geometrically representing a
gravitational field, such that the field equations in each representation
result equivalent to Einstein's field equations. Explicitly we mean by this
statement the following: any model of a gravitational field in GR represented
by\ a \textit{Lorentzian spacetime }(\textit{with non null Riemann curvature
tensor and null torsion tensor }which is also parallelizable\footnote{A
manifold $M$ is said to be parallelizable if it admits four \textit{global}
linearly independent vector fields.}) is equivalent to a teleparallel
spacetime (i.e., a spacetime structure equipped with a metrical compatible
teleparallel connection, which has \textit{null }Riemann curvature tensor and
\textit{non null }torsion tensor)\footnote{There are hundreds of papers (as
e.g., \cite{deandrade}) on the subject.} or equivalent to a special spacetime
structure, where the manifold $M$ is equipped with a Minkowski metric, and
where there is also defined a connection such that its \textit{nonmetricity
tensor} is not null. The teleparallel possibility is described in details in
Section 2 using the modern theory of differential forms and we claim that our
presentation leaves also clear that we can even dispense with the concept of a
\textit{connection} in the description of a gravitational
field\footnote{Explicitly, we mean that the gravitational field may be
interpreted as a field in the sense of Faraday, as it is the case of the
electromagnetic field.}, it is only necessary for such a representation to
exists that the manifold $M$ representing the set of all possible events be
parallelizable, admitting four global (not all exact) $1$-form fields coupled
in a specific way (see below). The second possibility is illustrated with an
example in Section 3. In Section 4 we present the conclusions.

\section{ Torsion as\ a Description of Gravity}

\subsection{Some Notation}

Suppose that a $4$-dimensional $M$ manifold is parallelizable, thus admitting
a set of four global linearly independent vector $%
\sle
_{\mathbf{a}}\in\sec TM$, $\mathbf{a}=0,1,2,3$ fields\footnote{We recall that
$\sec TM$ means section of the tangent bundle and $\sec T^{\ast}M$ means
section of the cotangent bundle. Also $\sec T_{s}^{r}M$ means the bundle of
tensors of type $(r,s)$ and $\sec%
{\displaystyle\bigwedge\nolimits^{r}}
T^{\ast}M$ a section of the bundle of $r$-forms fields.} such $\ \{%
\sle
_{\mathbf{a}}\}$ is a basis for $TM$ and let $\{\theta^{\mathbf{a}}%
\},\theta^{\mathbf{a}}\in\sec T^{\ast}M$ be the corresponding dual basis
($\theta^{\mathbf{a}}(%
\sle
_{\mathbf{b}})=\delta_{\mathbf{b}}^{\mathbf{a}}$). Suppose also that not all
the $\theta^{\mathbf{a}}$ are closed, i.e.,
\begin{equation}
d\theta^{\mathbf{a}}\neq0, \label{0}%
\end{equation}
for a least some $\mathbf{a}=0,1,2,3$. The $4$-form field $\theta^{\mathbf{0}%
}\wedge\theta^{\mathbf{1}}\wedge\theta^{\mathbf{2}}\wedge\theta^{\mathbf{3}}$
defines a (positive) orientation for $M$.

Now, the $\{\theta^{\mathbf{a}}\}$ can be used to define a Lorentzian metric
field in $M$ by defining $%
\slg
\in\sec T_{2}^{0}M$ by
\begin{equation}%
\slg
:=\eta_{\mathbf{ab}}\theta^{\mathbf{a}}\otimes\theta^{\mathbf{b}}, \label{1}%
\end{equation}
with the matrix with entries $\eta_{\mathbf{ab}}$ being the diagonal matrix
$(1,-1,-1,-1)$. Then, according to $%
\slg
$ the $\{\mathbf{e}_{\mathbf{a}}\}$ are orthonormal, i.e.,
\begin{equation}%
\sle
_{\mathbf{a}}\underset{%
\sslg
}{\cdot}%
\sle
_{\mathbf{b}}:=%
\slg
(%
\sle
_{\mathbf{a}},%
\sle
_{\mathbf{b}})=\eta_{\mathbf{ab}}. \label{2}%
\end{equation}

\begin{remark}
Since according to \emph{Eq.(\ref{2})} $%
\sle
_{\mathbf{0}}$ is a global time like vector field it follows it defines a time
orientation in $M$ which we denote by $\uparrow$. It follows that that the
4-tuple $(M,%
\slg
,\tau_{%
\sslg
},\uparrow)$ is part of a structure defining a \textit{Lorentzian spacetime}
and can serve as a \textit{substructure} to model a gravitational field in GR.
\end{remark}

For future use we also introduce $\mathtt{g}\in\sec T_{0}^{2}M$ by%
\begin{equation}
\mathtt{g}:=\eta^{\mathbf{ab}}%
\sle
_{\mathbf{a}}\otimes%
\sle
_{\mathbf{b}}, \label{3}%
\end{equation}
and we write:%
\begin{equation}
\theta^{\mathbf{a}}\underset{\mathtt{g}}{\cdot}\theta^{\mathbf{b}}%
:=\mathtt{g}(\theta^{\mathbf{a}},\theta^{\mathbf{b}})=\eta^{\mathbf{ab}}.
\label{4}%
\end{equation}

Due to the hypothesis given by Eq.(\ref{0}) the vector fields $\mathbf{e}%
_{\mathbf{a}}$, $\mathbf{a}=0,1,2,3$ will in general satisfy%
\begin{equation}
\lbrack%
\sle
_{\mathbf{a}},%
\sle
_{\mathbf{b}}]=c_{\mathbf{ab}}^{\mathbf{k}}%
\sle
_{\mathbf{k}}, \label{5}%
\end{equation}
where the $c_{\mathbf{ab}}^{\mathbf{k}}$ are the structure coefficients of the
basis $\{%
\sle
_{\mathbf{a}}\}$. It can been easily shown that \footnote{See,
e.g.,\cite{rodcap2007}.}%
\begin{equation}
d\theta^{\mathbf{a}}=-\frac{1}{2}c_{\mathbf{kl}}^{\mathbf{a}}\theta
^{\mathbf{k}}\wedge\theta^{\mathbf{l}}. \label{6}%
\end{equation}

Now, we introduce two different metric compatible connections on $M$, namely
$D$ (the Levi-Civita connection of $%
\slg
$) and a \textit{teleparallel} connection $\nabla$, such that%
\begin{align}
D_{%
\sle
_{\mathbf{a}}}%
\sle
_{\mathbf{b}}  &  =\omega_{\mathbf{ab}}^{\mathbf{c}}%
\sle
_{\mathbf{c}},\text{ }D_{%
\sle
_{\mathbf{a}}}\theta^{\mathbf{b}}=-\omega_{\mathbf{ac}}^{\mathbf{b}}%
\theta^{\mathbf{c}},\nonumber\\
\nabla_{%
\sle
_{\mathbf{a}}}%
\sle
_{\mathbf{b}}  &  =0,\text{ \ \ \ \ \ \ \ }\nabla_{%
\sle
_{\mathbf{a}}}\theta^{\mathbf{b}}=0. \label{7}%
\end{align}

The objects $\omega_{\mathbf{ab}}^{\mathbf{c}}$are called the connection
coefficients of the connection $D$ in the $\{%
\sle
_{\mathbf{a}}\}$ basis and the objects $\omega_{\mathbf{b}}^{\mathbf{a}}%
\in\sec T^{\ast}M$ defined by
\begin{equation}
\omega_{\mathbf{b}}^{\mathbf{a}}:=\omega_{\mathbf{kb}}^{\mathbf{a}}%
\theta^{\mathbf{k}}, \label{8}%
\end{equation}
are called the connection $1$-forms in the $\{%
\sle
_{\mathbf{a}}\}$ basis.

\begin{remark}
The connection coefficients $\varpi_{\mathbf{ac}}^{\mathbf{b}}$ of $\nabla$
and the connection $1$-forms of $\nabla$ in the basis $\{%
\sle
_{\mathbf{a}}\}$ are null according to the second line of \emph{Eq.(\ref{7})
and thus the basis }$\{%
\sle
_{\mathbf{a}}\}$ is called teleparallel. So, the connection $\nabla$ defines
an absolute parallelism on $M$. We recall that as said in the introduction
that idea has been introduced by Ricci.
\end{remark}

\begin{remark}
Of course, as it is well known the Riemann curvature tensor of $D$,\ is in
general non null in all points of $M$, but the torsion tensor of $D$ is zero
in all points of $M$. On the other hand the Riemann curvature tensor of
$\nabla$ is null in all points of $M$, whereas the torsion tensor of $\nabla$
is non null in all points of $M$.
\end{remark}

We recall also in order to fix notation that for a general connection, say
$\mathbf{D}$ on $M$ (not necessarily metric compatible) the \textit{torsion\/
and curvature operations} and the torsion and \textit{curvature \/}tensors of
a given general connection, say $\mathbf{D}$, are respectively the mappings:%
\begin{align}
\mathbf{\tau}(\mathbf{u,v})  &  =\mathbf{D}_{\mathbf{u}}\mathbf{v}%
-\mathbf{D}_{\mathbf{v}}\mathbf{u}-[\mathbf{u,v}],\label{top}\\
\mathbf{\rho(u,v)}  &  =\mathbf{D}_{\mathbf{u}}\mathbf{D}_{\mathbf{v}%
}-\mathbf{D}_{\mathbf{v}}\mathbf{D}_{\mathbf{u}}-\mathbf{D}_{[\mathbf{u,v}]}
\label{cop}%
\end{align}

and%
\begin{align}
\mathcal{T}(\alpha,\mathbf{u,v})  &  =\alpha\left(  \mathbf{\tau}(u,v)\right)
,\label{to op}\\
\mathbf{R}(\mathbf{w},\alpha,\mathbf{u,v})  &  =\alpha(\mathbf{\rho(u,v)w}),
\label{curv op}%
\end{align}
for every $\mathbf{u,v,w}\in\sec TM$ and $\alpha\in\sec\bigwedge^{1}T^{\ast}%
M$. In particular we write \
\begin{subequations}
\label{toricomp}%
\begin{align}
T_{\mathbf{bc}}^{\mathbf{a}}  &  :=\mathcal{T}(\theta^{\mathbf{a}},%
\sle
_{\mathbf{b}}\mathbf{,}%
\sle
_{\mathbf{c}})\label{ricomp}\\
R_{\mathbf{a\;cd}}^{\;\mathbf{b}}  &  :=\mathbf{R}(%
\sle
_{\mathbf{a}},\theta^{\mathbf{b}},%
\sle
_{\mathbf{c}},%
\sle
_{\mathbf{d}}),
\end{align}
and define the Ricci tensor by
\end{subequations}
\begin{subequations}
\begin{align}
Ricci  &  :=R_{\mathbf{ac}}\theta^{\mathbf{a}}\otimes\theta^{\mathbf{c}%
},\label{ric}\\
R_{\mathbf{ac}}  &  :=R_{\mathbf{a\;cb}}^{\;\mathbf{b}}=R_{\mathbf{ca}}.
\end{align}

We shall need also in order to fix our conventions to briefly recall the
definitions of the scalar product and left and right contractions on the so
called Hodge bundle $(%
{\displaystyle\bigwedge}
T^{\ast}M,%
\slg
)$ where $%
{\displaystyle\bigwedge}
T^{\ast}M=%
{\displaystyle\bigoplus\nolimits_{r=0}^{4}}
{\displaystyle\bigwedge\nolimits^{r}}
T^{\ast}M$ is the bundle of \textit{nonhomogeneous} multiforms.

So, if $\mathcal{A},\mathcal{B}\in\sec\bigwedge^{r}T^{\ast}M$ , $\mathcal{A}%
=\frac{1}{r!}A_{\mathbf{i}_{1}...\mathbf{i}_{\mathbf{r}}}\theta^{\mathbf{i}%
_{1}}\wedge...\wedge\theta^{\mathbf{i}_{\mathbf{r}}}$, $\mathcal{B}=\frac
{1}{r!}B_{\mathbf{j}_{1}...\mathbf{j}_{\mathbf{r}}}\theta^{\mathbf{j}_{1}%
}\wedge...\wedge\theta^{\mathbf{j}_{\mathbf{r}}}$, their scalar product
$\underset{g}{\cdot}$ (induced by $g$) is the \textit{linear} mapping%
\end{subequations}
\begin{equation}
\mathcal{A}\underset{g}{\cdot}\mathcal{B}=(\frac{1}{r!})A_{\mathbf{i}%
_{1}...\mathbf{i}_{\mathbf{r}}}B_{\mathbf{j}_{1}...\mathbf{j}_{\mathbf{r}}%
}\left(
\begin{array}
[c]{cccc}%
\theta^{\mathbf{i}_{1}}\underset{g}{\cdot}\theta^{\mathbf{j}_{1}} & ... &
... & \theta^{\mathbf{i}_{1}}\underset{g}{\cdot}\theta^{\mathbf{j}%
_{\mathbf{r}}}\\
... & ... & ... & ...\\
... & ... & ... & ...\\
\theta^{\mathbf{i}_{\mathbf{r}}}\underset{g}{\cdot}\theta^{\mathbf{j}_{1}} &
... & ... & \theta^{\mathbf{i}_{\mathbf{r}}}\underset{g}{\cdot}\theta
^{\mathbf{j}_{\mathbf{r}}}%
\end{array}
\right)  .\label{scp}%
\end{equation}
Also, for $\mathcal{A}\in\sec\bigwedge^{r}T^{\ast}M$, $\mathcal{C}\in
\sec\bigwedge^{s}T^{\ast}M$ it is $\mathcal{A}\underset{%
\sslg
}{\cdot}\mathcal{C}=0$. The left and right contractions of $\mathcal{X}%
,\mathcal{Y}$ $\in\sec%
{\displaystyle\bigwedge}
T^{\ast}M$ are defined for arbitrary (nonhomogeneous) multiforms as the
mappings $\underset{g}{\lrcorner}:%
{\displaystyle\bigwedge}
T^{\ast}M\times%
{\displaystyle\bigwedge}
T^{\ast}M\rightarrow%
{\displaystyle\bigwedge}
T^{\ast}M$ and $\underset{g}{\llcorner}:%
{\displaystyle\bigwedge}
T^{\ast}M\times%
{\displaystyle\bigwedge}
T^{\ast}M\rightarrow%
{\displaystyle\bigwedge}
T^{\ast}M$ such that for all $\mathcal{Z}$ $\in\sec%
{\displaystyle\bigwedge}
T^{\ast}M$.%
\begin{subequations}
\begin{align}
(\mathcal{X}\underset{g}{\lrcorner}\mathcal{Y)}\underset{g}{\cdot}\mathcal{Z}
&  =\mathcal{Y}\underset{g}{\cdot}(\mathcal{\tilde{X}}\wedge\mathcal{Z)}%
,\nonumber\\
(\mathcal{X}\underset{g}{\llcorner}\mathcal{Y)}\underset{g}{\cdot}\mathcal{Z}
&  =\mathcal{X}\underset{g}{\cdot}(Z\wedge\mathcal{\tilde{Y})},\label{con}%
\end{align}
where the \textit{tilde} means the operation of reversion, e.g., if
$\mathcal{B}=\frac{1}{r!}B_{\mathbf{i}_{1}...\mathbf{i}_{\mathbf{r}}}%
\theta^{\mathbf{i}_{1}}\wedge...\wedge\theta^{\mathbf{i}_{\mathbf{r}}}$, then
$\mathcal{\tilde{B}}=\frac{1}{r!}B_{\mathbf{i}_{1}...\mathbf{i}_{\mathbf{r}}%
}\theta^{\mathbf{i}_{\mathbf{r}}}\wedge...\wedge\theta^{\mathbf{i}_{1}}$.

The \textit{Hodge star operator\/} (or Hodge dual) is the linear mapping
$\underset{g}{\star}:\bigwedge^{r}T^{\ast}M\rightarrow\bigwedge^{n-r}T^{\ast
}M$
\end{subequations}
\begin{equation}
\mathcal{A}\wedge\underset{g}{\star}\mathcal{B}=(\mathcal{A}\underset{g}%
{\cdot}\mathcal{B})\tau_{%
\sslg
}, \label{star}%
\end{equation}
\ for every $\mathcal{A},\mathcal{B}\in\bigwedge^{r}T^{\ast}M$. The inverse
$\underset{g}{\star}^{-1}:\bigwedge^{n-r}T^{\ast}M\rightarrow\bigwedge
^{r}T^{\ast}M$ of the Hodge star operator is given by%
\begin{equation}
\underset{g}{\star}^{-1}=(-1)^{r(n-r)}\mathrm{sgn}\overset{.}{%
\slg
}\underset{g}{\star}, \label{star1}%
\end{equation}
where \textrm{sgn }$\overset{}{%
\slg
}$ $=\det\overset{}{%
\slg
}/|\det\overset{}{%
\slg
}|$ denotes the sign of the determinant \ of the matrix with entries
$g_{ij}=\overset{}{%
\slg
}(%
\sle
_{i},%
\sle
_{j})$. The \textit{Hodge coderivative operator} $\underset{g}{\delta}$
(associated to $%
\slg
$) is defined for $\mathcal{A}\in\sec\bigwedge^{r}T^{\ast}M$ by%
\begin{equation}
\underset{g}{\delta}\mathcal{A}=(-1)^{r}\underset{g}{\star}^{-1}%
d\;\underset{g}{\star}\mathcal{A}. \label{hodgecod}%
\end{equation}

\subsection{Cartan's Structure Equations}

Given that we introduced two different connections $D$ and $\nabla$ defined in
the manifold $M$ we can write \textit{two different pairs} of Cartan's
structure equations, each one of the pairs describing respectively the
geometry of the structures $(M,D,%
\slg
,\tau_{%
\sslg
},\uparrow)$ and $(M,\nabla,%
\slg
,\tau_{%
\sslg
},\uparrow)$ which will be called respectively a Lorentzian spacetime and a
teleparallel spacetime.

\subsubsection{Cartan's Structure Equations for $D$}

In this case we write
\begin{align}
\Theta^{\mathbf{a}}  &  :=d\theta^{a}+\omega_{\mathbf{b}}^{\mathbf{a}}%
\wedge\theta^{\mathbf{b}}=0,\nonumber\\
\mathcal{R}_{\mathbf{b}}^{\mathbf{a}}  &  :=d\omega_{\mathbf{b}}^{\mathbf{a}%
}+\omega_{\mathbf{c}}^{\mathbf{a}}\wedge\omega_{\mathbf{b}}^{\mathbf{c}},
\label{9}%
\end{align}
where the $\Theta^{\mathbf{a}}\in\sec%
{\displaystyle\bigwedge\nolimits^{2}}
T^{\ast}M$, $\mathbf{a}=0,1,2,3$ and the $\mathcal{R}_{\mathbf{b}}%
^{\mathbf{a}}\in\sec%
{\displaystyle\bigwedge\nolimits^{2}}
T^{\ast}M$, $\mathbf{a},\mathbf{b}=0,1,2,3$ are respectively the torsion and
the curvature $\ 2$-forms of $D$ with%
\begin{equation}
\Theta^{\mathbf{a}}=\frac{1}{2}T_{\mathbf{bc}}^{\mathbf{a}}\theta^{\mathbf{b}%
}\wedge\theta^{\mathbf{c}},\text{ }\mathcal{R}_{\mathbf{b}}^{\mathbf{a}}%
=\frac{1}{2}R_{\mathbf{b\;cd}}^{\;\mathbf{a}}\theta^{\mathbf{c}}\wedge
\theta^{\mathbf{d}}\text{.} \label{trc}%
\end{equation}

\subsubsection{Cartan's Structure Equations for $\nabla$}

In this case since $\varpi_{\mathbf{b}}^{\mathbf{a}}=0$ we have
\begin{align}
\bar{\Theta}^{\mathbf{a}}  &  :=d\theta^{a}+\varpi_{\mathbf{b}}^{\mathbf{a}%
}\wedge\theta^{\mathbf{b}}=d\theta^{a},\nonumber\\
\mathcal{\bar{R}}_{\mathbf{b}}^{\mathbf{a}}  &  :=d\varpi_{\mathbf{b}%
}^{\mathbf{a}}+\varpi_{\mathbf{c}}^{\mathbf{a}}\wedge\varpi_{\mathbf{b}%
}^{\mathbf{c}}=0, \label{10}%
\end{align}
where the $\bar{\Theta}^{\mathbf{a}}\in\sec%
{\displaystyle\bigwedge\nolimits^{2}}
T^{\ast}M$, $\mathbf{a}=0,1,2,3$ and the $\mathcal{\bar{R}}_{\mathbf{b}%
}^{\mathbf{a}}\in\sec%
{\displaystyle\bigwedge\nolimits^{2}}
T^{\ast}M$, $\mathbf{a},\mathbf{b}=0,1,2,3$ \ and are respectively the torsion
and the curvature $\ 2$-forms of $\nabla$ given by formulas analogous to the
ones in Eq.(\ref{trc}).

We next suppose that the $\{\theta^{\mathbf{a}}\}$ are the basic variables
representing a gravitation field. We postulate for the $\{\theta^{a}\}$
interacts with the matter fields though the following Lagrangian
density\footnote{We observe that the first term in Eq.(\ref{11}) is just the
Lagrangian density used by Einstein in \cite{einstein4}.}%
\begin{equation}
\mathcal{L=L}_{g}+\mathcal{L}_{m}, \label{10b}%
\end{equation}
where $\mathcal{L}_{m}$ is the matter Lagrangian density and%

\begin{equation}
\mathcal{L}_{g}=-\frac{1}{2}d\theta^{\mathbf{a}}\wedge\underset{g}{\star
}d\theta_{\mathbf{a}}+\frac{1}{2}\underset{g}{\delta}\theta^{\mathbf{a}}%
\wedge\underset{g}{\star}\underset{g}{\delta}\theta_{\mathbf{a}}+\frac{1}%
{4}\left(  d\theta^{\mathbf{a}}\wedge\theta_{\mathbf{a}}\right)
\wedge\underset{g}{\star}\left(  d\theta^{\mathbf{b}}\wedge\theta_{\mathbf{b}%
}\right)  , \label{11}%
\end{equation}

\begin{remark}
This Lagrangian is not invariant under arbitrary point dependent Lorentz
rotations of the basic cotetrad fields. In fact, if $\theta^{\mathbf{a}%
}\mapsto\theta^{\prime\mathbf{a}}=\Lambda_{\mathbf{b}}^{\mathbf{a}}%
\theta^{\mathbf{b}}$, where for each $x\in M$, $\Lambda_{\mathbf{b}%
}^{\mathbf{a}}(x)\in L_{+}^{\uparrow}$ \emph{(}the homogeneous
and\ orthochronous Lorentz group\emph{) we get that}
\end{remark}

\begin{equation}
\mathcal{L}_{g}^{\prime}=-\frac{1}{2}d\theta^{\prime\mathbf{a}}\wedge
\underset{g}{\star}d\theta_{\mathbf{a}}^{\prime}+\frac{1}{2}\underset
{g}{\delta}\theta^{\prime\mathbf{a}}\wedge\underset{g}{\star}\underset
{g}{\delta}\theta_{\mathbf{a}}^{\prime}+\frac{1}{4}\left(  d\theta
^{\prime\mathbf{a}}\wedge\theta_{\mathbf{a}}^{\prime}\right)  \wedge
\underset{g}{\star}\left(  d\theta^{\prime\mathbf{b}}\wedge\theta_{\mathbf{b}%
}^{\prime}\right)  ,
\end{equation}
differs from $\mathcal{L}_{g}$ by an exact differential. So, the field
equations derived by the variational principle results invariant under a
change of gauge.\footnote{See details in \cite{rodcap2007}.}

Now, to derive the field equations directly from Eq.(\ref{11}) is a nontrivial
and laborious exercise, whose details the interested reader may find in
\cite{rodcap2007}. The result is:%

\begin{equation}
d\underset{g}{\star}\mathcal{S}_{\mathbf{d}}+\underset{g}{\star}t_{\mathbf{d}%
}=-\underset{g}{\star}\mathcal{T}_{\mathbf{d}}, \label{12}%
\end{equation}
where%
\begin{align}
\underset{g}{\star}t_{\mathbf{d}}  &  :=\frac{\partial\mathcal{L}_{g}%
}{\partial\theta^{\mathbf{d}}}=\frac{1}{2}[(\theta_{\mathbf{d}}\underset
{g}{\lrcorner}d\theta^{\mathbf{a}})\wedge\underset{g}{\star}d\theta
_{\mathbf{a}}-d\theta^{\mathbf{a}}\wedge(\theta_{\mathbf{d}}\underset
{g}{\lrcorner}\underset{g}{\star}d\theta_{\mathbf{a}})]\nonumber\\
&  +\frac{1}{2}d\left(  \theta_{\mathbf{d}}\underset{g}{\lrcorner}\underset
{g}{\star}\theta^{\mathbf{a}}\right)  \wedge\underset{%
\sslg
}{\star}d\underset{%
\sslg
}{\star}\theta_{\mathbf{a}}+\frac{1}{2}d\left(  \theta_{\mathbf{d}}%
\underset{g}{\lrcorner}\underset{g}{\star}\theta^{\mathbf{a}}\right)
\wedge\underset{g}{\star}d\underset{g}{\star}\theta_{\mathbf{a}}+\frac{1}%
{2}d\theta_{\mathbf{d}}\wedge\underset{g}{\star}\left(  d\theta^{\mathbf{a}%
}\wedge\theta_{\mathbf{a}}\right) \nonumber\\
&  -\frac{1}{4}d\theta^{\mathbf{a}}\wedge\theta_{\mathbf{a}}\wedge\left[
\theta_{\mathbf{d}}\underset{g}{\lrcorner}\underset{g}{\star}\left(
d\theta^{\mathbf{c}}\wedge\theta_{\mathbf{c}}\right)  \right]  -\frac{1}%
{4}\left[  \theta_{\mathbf{d}}\underset{g}{\lrcorner}\left(  d\theta
^{\mathbf{c}}\wedge\theta_{\mathbf{c}}\right)  \right]  \wedge\underset
{g}{\star}\left(  d\theta^{\mathbf{a}}\wedge\theta_{\mathbf{a}}\right)  ,
\label{13}%
\end{align}%
\begin{equation}
\underset{g}{\star}\mathcal{S}_{\mathbf{d}}:=\frac{\partial\mathcal{L}_{g}%
}{\partial d\theta^{\mathbf{d}}}=-\underset{g}{\star}d\theta_{\mathbf{d}%
}-\left(  \theta_{\mathbf{d}}\underset{g}{\lrcorner}\underset{g}{\star}%
\theta^{\mathbf{a}}\right)  \wedge\underset{g}{\star}d\underset{g}{\star
}\theta_{\mathbf{a}}+\frac{1}{2}\theta_{\mathbf{d}}\wedge\underset{g}{\star
}\left(  d\theta^{\mathbf{a}}\wedge\theta_{\mathbf{a}}\right)  . \label{14}%
\end{equation}
and the\footnote{We suppose that $\mathcal{L}_{m}$ does not depend explicitly
on the $d\theta^{\mathbf{a}}$.} \
\begin{equation}
\underset{g}{\star}\mathcal{T}_{\mathbf{d}}:=\frac{\partial\mathcal{L}_{m}%
}{\partial\theta^{\mathbf{d}}}=-\underset{g}{\star}T_{\mathbf{d}} \label{15}%
\end{equation}
are the energy-momentum $3$-forms of the matter fields\footnote{In reality,
due the conventions used in this paper the true energy-momentum $3$-forms are
$\underset{g}{\star}T_{\mathbf{d}}=-\underset{g}{\star}\mathcal{T}%
_{\mathbf{d}}$.}.

Recalling \ that from Eq.(\ref{10}) \ it is $\bar{\Theta}^{\mathbf{a}%
}:=d\theta^{a}$, the field equations (Eq.(\ref{12})) can be written as%
\begin{equation}
d\underset{g}{\star}\mathcal{F}_{\mathbf{d}}=-\underset{g}{\star}%
\mathcal{T}_{\mathbf{d}}-\underset{g}{\star}t_{\mathbf{d}}-\underset{g}{\star
}\mathfrak{h}_{d}, \label{16}%
\end{equation}

where%
\begin{equation}
\mathfrak{h}_{\mathbf{d}}=d\left[  \left(  \theta_{\mathbf{d}}\underset
{g}{\lrcorner}\underset{g}{\star}\theta^{\mathbf{a}}\right)  \wedge
\underset{g}{\star}d\underset{g}{\star}\theta_{\mathbf{a}}-\frac{1}{2}%
\theta_{\mathbf{d}}\wedge\underset{g}{\star}\left(  d\theta^{\mathbf{a}}%
\wedge\theta_{\mathbf{a}}\right)  \right]  . \label{17}%
\end{equation}

Finally recalling the definition of the Hodge coderivative operator ( Eq.(
\ref{hodgecod})) we can write Eq.(\ref{16}) as
\begin{equation}
\underset{g}{\delta}\mathcal{F}^{\mathbf{d}}=-(\mathcal{T}_{\mathbf{\ }%
}^{\mathbf{d}}+\mathbf{t}^{\mathbf{d}}), \label{19}%
\end{equation}
with the $\mathbf{t}^{\mathbf{d}}\in\sec%
{\displaystyle\bigwedge\nolimits^{1}}
T^{\ast}M$ given by
\begin{equation}
\mathbf{t}^{\mathbf{d}}:=t^{\mathbf{d}}+\mathfrak{h}^{\mathbf{d}}, \label{20}%
\end{equation}
legitimate energy-momenta\footnote{Indeed, for each index$\ \mathbf{d}$ the
first member of Eq.(\ref{19}) is a $1$-from field and also $\mathcal{T}%
_{\mathbf{\ }}^{\mathbf{d}}$ is an $1$-form field , so $\delta\mathcal{F}%
^{\mathbf{d}}+\mathcal{T}_{\mathbf{\ }}^{\mathbf{d}}=-\mathbf{t}^{\mathbf{d}}$
is a $1$-form field .} $1$-form fields for the gravitational field. Note that
the total energy momentum tensor of matter plus the gravitational field is
trivially conserved in our theory, i.e.,
\begin{equation}
\underset{g}{\delta}(\mathcal{T}_{\mathbf{\ }}^{\mathbf{d}}+\mathbf{t}%
^{\mathbf{d}})=0, \label{21}%
\end{equation}
since $\underset{g}{\delta}^{2}\mathcal{F}^{\mathbf{d}}=0$.

\begin{remark}
In \emph{\cite{notterod}} a theory of the gravitational field in Minkowski
spacetime $(M\simeq\mathbb{R}^{4},\overset{\circ}{%
\slg
},D,\tau_{\overset{\circ}{%
\sslg
}},\uparrow)$ has been presented where a nontrivial gravitational field
configuration was interpreted as generating an effective Lorentzian spacetime
$(M\simeq\mathbb{R}^{4},%
\slg
,D,\tau_{g},\uparrow)$ where $%
\slg
$ satisfies Einstein equations and where probe particles and/or fields move.
It was assumed there that the gravitational field $%
\slg
=\eta_{\mathbf{ab}}\theta^{\mathbf{a}}\otimes\theta^{\mathbf{b}}$ is a field
in Faraday sense\footnote{This means that it is interpreted as a field with an
ontology \ analogous to \ the electromagnetic field.}, i.e., the fields
$\theta^{\mathbf{a}}$ have their dynamics described by a \emph{(}%
postulated\emph{)} Lagrangian density like the one in \emph{Eq.(\ref{11}).
}Moreover, it was postulated that the $\theta^{\mathbf{a}}$ couple universally
with the matter fields and that the presence of energy-momentum due to matter
fields in some region of Minkowski spacetime distorts the Lorentz vacuum in
much the same way that stresses in an elastic body distorts it. Now,
distortions \emph{(}or deformations\emph{)} in the theory of dislocations
according to \emph{\cite{Zorawski}} can be of the \textit{elastic} or
\textit{plastic} type. An elastic distortion is described by a diffeomorphism
\texttt{h}$:M\rightarrow M$. In this case the induced metric
is$\ \mathtt{\mathbf{g}}=\mathtt{h}^{\ast}%
\mbox{\boldmath{$\eta$}}%
$ (analogous to the Cauchy-Green tensor \emph{\cite{frankel}} of elasticity
theory) and according to \emph{Remark 250}\ in \emph{\cite{rodcap2007}} its
Levi-Civita connection is $\mathtt{h}^{\ast}\dot{D}$. This implies that the
structure $(\mathtt{h}M\simeq\mathbb{R}^{4},\mathtt{\mathbf{g}},\mathtt{h}%
^{\ast}\mathring{D},\tau_{g},\uparrow)$ is again Minkowski spacetime. In the
original versions of \emph{\cite{notterod, rodcap2007}} this was the type of
deformation considered, but this has been corrected in improved versions of
those manuscripts, respectively at the arXiv and at
http://ime.unicamp.br/walrod/recentes where an errata to \emph{Chapter 10 of}
\emph{\cite{rodcap2007}} at . In the quoted errata, the deformation is taken
to be of the plastic type and represented by a distortion extensor field $M$,
i.e., a linear mapping $h:\bigwedge\nolimits^{1}T^{\ast}M\rightarrow
\bigwedge\nolimits^{1}T^{\ast}M$ such that $g=h^{\dagger}h,$ where $g$ is the
extensor field associated with $%
\slg
$ \emph{(}see below\emph{)\footnote{Recently we take knowledge that Coll
\cite{coll} found that any Lorentzian metric can be written as a deformation
of the Minkowski metric involving a $2$-form field. We will investigate in
another publication the relationship of ours and Coll's ideas.}.} A paper
describing a detailed theory of gravitation as a plastic deformation of the
Lorentz vacuum is in preparation \emph{\cite{fr}}.
\end{remark}

\begin{remark}
In electromagnetic theory on a Lorentzian spacetime we have only one potential
$A\in\sec%
{\displaystyle\bigwedge\nolimits^{1}}
T^{\ast}M$ and the field equations are%
\begin{equation}
dF=0,\text{ }\underset{g}{\delta}F=-J,\text{ } \label{Maxwell}%
\end{equation}
where $F\in\sec%
{\displaystyle\bigwedge\nolimits^{2}}
T^{\ast}M$ is the electromagnetic field and $J\in\sec%
{\displaystyle\bigwedge\nolimits^{1}}
T^{\ast}M$ $\ $is the electric current. The two Maxwell equations in
\emph{Eq.(\ref{Maxwell})} can be written as a single equation using the
Clifford bundle formalism \cite{rodcap2007}. In this formalism $%
{\displaystyle\bigwedge}
T^{\ast}M\hookrightarrow\mathcal{C\ell(}M,g)$ . Then it can be shown that in
this case $\partial=d-\underset{g}{\delta}=\theta^{\mathbf{a}}D_{\mathbf{e}%
_{\mathbf{a}}}$ is the Dirac operator\emph{ (}acting on sections of
$\mathcal{C\ell(}M,g)$\emph{)} and we can write Maxwell equation as%
\begin{equation}
\partial F=J. \label{ma}%
\end{equation}
Now, if you fell uncomfortable in needing four distinct potentials
$\theta^{\mathbf{a}}$ for describing the gravitational field you can put then
together defining a vector valued differential form%
\begin{equation}
\theta=\theta^{\mathbf{a}}\otimes%
\sle
_{\mathbf{a}}, \label{gpot}%
\end{equation}
and in this case the gravitational field equations are%
\begin{align}
d\bar{\Theta}  &  =0,\nonumber\\
\underset{g}{\delta}\bar{\Theta}  &  =-(\mathcal{T}_{\mathbf{\ }}+\mathbf{t}),
\label{g2}%
\end{align}
where $\bar{\Theta}\mathcal{=}\bar{\Theta}^{\mathbf{a}}\otimes%
\sle
_{\mathbf{a}},\mathcal{T}=\mathcal{T}^{\mathbf{a}}\otimes%
\sle
_{\mathbf{a}},\mathbf{t=t}^{\mathbf{a}}\otimes%
\sle
_{\mathbf{a}}$. By considering the bundle $\mathcal{C\ell(}M,g)\otimes TM$ we
can even write the two equations in \emph{Eq.(\ref{g2})} as a single equation,
i.e.,%
\begin{equation}
\partial\bar{\Theta}=\mathcal{T}_{\mathbf{\ }}+\mathbf{t} \label{g3}%
\end{equation}

\end{remark}

\subsection{Relation with Einstein's Theory}

At this point the reader may be asking: which is the relation of the theory
just presented with Einstein's GR theory? The answer is that recalling that
the connection $1$-forms $\omega^{\mathbf{cd}}$ of $D$ are given by
\begin{equation}
\omega^{\mathbf{cd}}=\frac{1}{2}\left[  \theta^{\mathbf{d}}\underset
{g}{\lrcorner}d\theta^{\mathbf{c}}-\theta^{\mathbf{c}}\underset{g}{\lrcorner
}d\theta^{\mathbf{d}}+\theta^{\mathbf{c}}\underset{g}{\lrcorner}\left(
\theta^{\mathbf{d}}\underset{g}{\lrcorner}d\theta_{\mathbf{a}}\right)
\theta^{\mathbf{a}}\right]  \label{22}%
\end{equation}
we can show through a laborious (but standard) exercise (see \cite{rodcap2007}
for details) that \ the first member of Eq.(\ref{12}) is exactly
$-\underset{g}{\star}\mathcal{G}_{\mathbf{d}}$ (the Einstein $3$-forms). So,
we have%
\begin{equation}
\underset{g}{\star}\mathcal{G}^{\mathbf{d}}:=\underset{g}{\star}%
(\mathcal{R}^{\mathbf{d}}-\frac{1}{2}R\theta^{\mathbf{d}}), \label{23}%
\end{equation}
with $\mathcal{R}^{\mathbf{d}}=R_{\mathbf{a}}^{\mathbf{d}}\theta
^{\mathbf{a}\text{ }}$the Ricci $1$-forms and $R$ the scalar curvature. Then
Eq.(\ref{12}) results equivalent to
\begin{equation}
\mathcal{R}^{\mathbf{d}}-\frac{1}{2}R\theta^{\mathbf{d}}=-T^{\mathbf{d}}
\label{24}%
\end{equation}
and taking the dot product of both members with $\theta_{\mathbf{a}}$ we get
\begin{equation}
R_{\mathbf{a}}^{\mathbf{d}}-\frac{1}{2}R\delta_{\mathbf{a}}^{\mathbf{d}%
}=-T_{\mathbf{a}}^{\mathbf{d}}, \label{25}%
\end{equation}
which is the usual tensorial form of Einstein's equations.

\begin{remark}
When the $\theta^{\mathbf{a}}$ and the $d\theta_{\mathbf{a}}$ are packed in
the form of the connection $1$-forms the Lagrangian density\ $\mathfrak{L}%
_{g\text{ }}$ becomes
\begin{equation}
\ \mathfrak{L}_{g\text{ }}=\mathcal{L}_{EH}+d(\theta^{\mathbf{a}}%
\wedge\underset{g}{\star}d\theta_{\mathbf{a}}), \label{26}%
\end{equation}
where
\begin{equation}
\mathcal{L}_{EH}=\frac{1}{2}\mathcal{R}_{\mathbf{cd}}\wedge\underset{g}{\star
}(\theta^{\mathbf{c}}\wedge\theta^{\mathbf{d}}) \label{27}%
\end{equation}
\emph{(}with $\mathcal{R}_{\mathbf{cd}}$ given by \emph{Eq.(\ref{9}))} is the
Einstein-Hilbert Lagrangian density.
\end{remark}

\section{A Comment on Einstein Most Happy Though}

The exercises presented above indicates that a geometrical interpretation for
the gravitational field is no more than an option among many ones. Indeed, it
is not necessary to introduce any connection $D$ \ or $\nabla$ on $M$ to have
a perfectly well defined theory of the gravitational field whose field
equations are equivalent to the Einstein field equations. Note that we have
not give until now any details on the \textit{global topology} of the world
manifold $M$. However, since we admitted that $M$ carries four global (not all
closed) $1$-form fields $\theta^{\mathbf{a}}$ which defines the object $%
\slg
$ it follows that $(M,D,%
\slg
,\tau_{%
\sslg
},\uparrow)$ is a spin manifold \cite{geroch,rodcap2007}, i.e., admit spinor
fields. This, of course, is necessary if the theory is to be useful in the
real world since fundamental matter fields are spinor fields. The most simple
spin manifold is clearly Minkowski spacetime which is represented by a
structure $(M,\mathring{D},\mathbf{\eta},\tau_{\eta},\uparrow)$ where
\ $M\simeq\mathbb{R}^{4\text{ }}$, and $\mathring{D}$ is the Levi-Civita
connection of the Minkowski metric $\mathbf{\eta}$. In that case it is
possible to interpret $%
\slg
$ as a field in the Faraday sense living in $(M,\mathring{D},\mathbf{\eta
},\tau_{\mathbf{\eta}},\uparrow)$, or to work directly with the $\theta
^{\mathbf{a}}$ which has a well defined dynamics and coupling to the matter fields.

At last we want to comment that as well known in Einstein's GR one can easily
distinguish in any \textit{real} \textit{physical laboratory} \cite{ohanian}
(despite some claims on the contrary) a true gravitational field from an
acceleration field of a given reference frame in Minkowski spacetime. This is
because in GR the \textit{mark} of a real gravitational field is the non null
Riemann curvature tensor of $D$, and the Riemann curvature tensor of the
Levi-Civita connection of $\mathring{D}$ (present in the definition of
Minkowski spacetime) is null. However if we interpret a gravitational field as
the torsion $2$-forms on the structure $(M,\nabla,%
\slg
,\tau_{%
\sslg
},\uparrow)$ viewed as a deformation of Minkowski spacetime then one can also
interpret an acceleration field of an accelerated reference frame in Minkowski
spacetime as generating an effective teleparallel spacetime $(M,\overset
{e}{\nabla},\eta,\tau_{\eta},\uparrow)$. This can be done as follows. Let
$Z\in\sec TU$, $U\subset M$ with $\mathbf{\eta}(Z,Z)=1$ an accelerated
reference frame on Minkowski spacetime. This means (see, e.g.,
\cite{rodcap2007} for details) that
\begin{equation}%
\sla
=\mathring{D}_{Z}Z=0. \label{28}%
\end{equation}
Call $e_{\mathbf{0}}=Z$ and define an accelerated reference frame as
\textit{non} trivial $\ $if $\vartheta^{\mathbf{0}}=\eta(Z,)$ is not an exact
differential. Next recall that in $U\subset M$ there always exist three other
$\mathbf{\eta}$-orthonormal vector fields $e_{\mathbf{i}}$, $\mathbf{i}=1,2,3$
such that $\{e_{a}\}$ is an $\mathbf{\eta}$-orthonormal \ basis for $TU$,
i.e.,%
\[
\mathbf{\eta}=\eta_{\mathbf{ab}}\vartheta^{\mathbf{a}}\otimes\vartheta
^{\mathbf{b}},
\]
where $\{\vartheta^{\mathbf{a}}\}$ be the dual basis\footnote{In general we
will also have that $d\vartheta^{\mathbf{i}}\neq0$, $\mathbf{i}=1,2,3$.} of
$\{e_{a}\}$. We then have
\begin{equation}
\mathring{D}_{e_{\mathbf{a}}}e_{\mathbf{b}}=\mathring{\omega}_{\mathbf{ab}%
}^{\mathbf{c}}e_{\mathbf{c}},\mathring{D}_{e_{\mathbf{a}}}\vartheta
^{\mathbf{b}}=-\mathring{\omega}_{\mathbf{ac}}^{\mathbf{b}}\vartheta
^{\mathbf{c}}. \label{29}%
\end{equation}

What remains in order to be possible to interpret an acceleration field as a
kind of\ `gravitational field' is to introduce on $M$ a $\eta$-metrical
compatible connection $\overset{e}{\nabla}$ such that the $\{e_{a}\}$ is
teleparallel according to it. We have%

\begin{equation}
\overset{e}{\nabla}_{e_{\mathbf{a}}}e_{\mathbf{b}}=0,\overset{e}{\nabla
}_{e_{\mathbf{a}}}\vartheta^{\mathbf{b}}=0. \label{30}%
\end{equation}

With this connection the structure $(M\simeq\mathbb{R}^{4},\overset{e}{\nabla
},\mathbf{\eta},\tau_{\eta},\uparrow)$ has null Riemann curvature tensor but a
non null torsion tensor, which an easy calculation shows to be related with
the acceleration and the other coefficients $\mathring{\omega}_{\mathbf{ab}%
}^{\mathbf{c}}$ of the connection $\mathring{D}$ in that basis, which describe
the motion on Minkowski spacetime of a \textit{grid} represented by the
orthonormal frame $\{e_{a}\}$. Sch\"{u}cking \cite{schu} \ thinks that such a
description of the gravitational field makes Einstein most happy though, i.e.,
the equivalence principle (understood as equivalence between acceleration and
gravitational field) a legitimate mathematical idea. However, a \textit{true}
gravitational field must satisfy (at least with good approximation)
Eq.(\ref{16}), whereas there is no single reason for an acceleration field to
satisfy that equation.

\section{ A Model for the Gravitational Field Represented by the Nonmetricity
of a Connection}

In this section we suppose that the world manifold $M$ is a $4$-dimensional
manifold diffeomorphic to $\mathbb{R}^{4}$. Let moreover $(t,x,y,z)=(x^{0}%
,x^{1},x^{2},x^{3})$\ be global Cartesian coordinates for $M$.

Next, introduce on $M$ two metric fields:%
\begin{equation}
\mathbf{\eta}=dt\otimes dt-dx^{1}\otimes dx^{1}-dx^{2}\otimes dx^{2}%
-dx^{3}\otimes dx^{3}, \label{1n}%
\end{equation}
and%
\begin{align}%
\slg
&  =\left(  1-\frac{2m}{r}\right)  dt\otimes dt\nonumber\\
&  -\left\{  \left(  1-\frac{2m}{r}\right)  ^{-1}-1\right\}  r^{-2}\left[
(x^{1})^{2}dx^{1}\otimes dx^{1}+(x^{2})^{2}dx^{2}\otimes dx^{2}+(x^{3}%
)^{2}dx^{3}\otimes dx^{3}\right] \nonumber\\
&  -dx^{1}\otimes dx^{1}+\left(  1-\frac{2m}{r}\right)  ^{-1}(dx^{2}\otimes
dx^{2}-dx^{3}\otimes dx^{3}). \label{2n}%
\end{align}
In Eq. (\ref{2n})
\begin{equation}
r=\sqrt{(x^{1})^{2}+(x^{2})^{2}+(x^{3})^{2}}. \label{4n}%
\end{equation}
\medskip

Now, introduce \ $(t,r,\vartheta,\varphi)=(x^{0},x^{\prime1},x^{\prime
2},x^{\prime3})$ as the usual spherical coordinates for $M$. \ Recall that
\begin{equation}
x^{1}=r\sin\vartheta\cos\varphi,\text{ }x^{2}=r\sin\vartheta\sin\varphi,\text{
}x^{3}=r\cos\vartheta\label{5n}%
\end{equation}
and the range of these coordinates in $\eta$ are $r>0$, $0<\vartheta<\pi$,
$0<\varphi<2\pi$. For $g$ the range of the $r$ variable must be $(0,2m)\cup
(2m,\infty)$.\medskip

\textbf{ }As can be easily verified, the metric $%
\slg
$ in spherical coordinates is:%
\begin{equation}%
\slg
=\left(  1-\frac{2m}{r}\right)  dt\otimes dt-\left(  1-\frac{2m}{r}\right)
^{-1}dr\otimes dr-r^{2}(d\vartheta\otimes d\vartheta+\sin^{2}\vartheta
d\varphi\otimes d\varphi), \label{6n}%
\end{equation}
which we immediately recognize as the \textit{Schwarzschild metric }of
GR\textit{.} Of course, $\mathbf{\eta}$ is a Minkowski metric on $M$.

As next step we introduce two \textit{distinct} connections, $\mathring{D}$
and $D$ on $M$. We assume that $\mathring{D}$ is the Levi-Civita connection of
$\mathbf{\eta}$ in $M$ and $D$ is the Levi-Civita connection of $%
\slg
$ in $M$. Then, by definition (see, e.g., \cite{rodcap2007} for more details)
the ammetricities tensors of $\mathring{D}$ relative to $\mathbf{\eta}$ and of
$D$ relative to $%
\slg
$ are null, i.e.,
\begin{align}
\mathring{D}\eta &  =0,\text{ }\nonumber\\
D%
\slg
&  =0. \label{7n}%
\end{align}

However, the nonmetricity tensor \ $\mathbf{A}^{\eta}\in\sec T_{3}^{0}M$ of
$\mathring{D}$ relative to $%
\slg
$ is non null, i.e.,
\begin{equation}
\mathring{D}%
\slg
=\mathbf{A}^{\eta}\mathbf{\neq}0, \label{8n}%
\end{equation}
and also the nonmetricity tensor \ $\mathbf{A}^{g}\in\sec T_{3}^{0}M$ of $D$
relative to $\mathbf{\eta}$ is non null, i.e.,%
\begin{equation}
D\mathbf{\eta}=\mathbf{A}^{g}\mathbf{\neq}0. \label{8nn}%
\end{equation}

We now calculate the components of $\mathbf{A}^{\eta}$ in the coordinated
bases $\{{\mbox{\boldmath$\partial$}}_{\mu}\}$ for $TM$ and $\{dx^{\nu}\}$ for
$T^{\ast}M$ associated with the coordinates $(x^{0},x^{1},x^{2},x^{3})$ of
$M$. Since $\mathring{D}$ is the Levi-Civita connection of the Minkowski
metric $\mathbf{\eta}$ we have that
\begin{align}
\mathring{D}_{\mbox{\tiny\boldmath$\partial$}_{\mu}}%
{\mbox{\boldmath$\partial$}}_{\nu}  &  =L_{\mu\nu}^{\rho}%
{\mbox{\boldmath$\partial$}}_{\rho}=0,\nonumber\\
\mathring{D}_{\mbox{\tiny\boldmath$\partial$}_{\mu}}dx^{\alpha}  &
=-L_{\mu\nu}^{\alpha}dx^{\nu}=0. \label{9n}%
\end{align}
i.e., the connection coefficients $L_{\mu\nu}^{\rho}$ of $\mathring{D}$ in
this basis are \textit{null. }Then, $\mathbf{A}^{\eta}=Q_{\mu\alpha\beta
}dx^{\alpha}\otimes dx^{\beta}\otimes dx^{\mu}$ is given by
\begin{align}
\mathbf{A}^{\eta}  &  =\mathring{D}%
\slg
=\mathring{D}_{\mbox{\tiny\boldmath$\partial$}_{\mu}}\left(  g_{\alpha\beta
}dx^{\alpha}\otimes dx^{\beta}\right)  \otimes dx^{\mu}\nonumber\\
&  =(\frac{\partial g_{\alpha\beta}}{\partial x^{\mu}})dx^{\alpha}\otimes
dx^{\beta}\otimes dx^{\mu}. \label{10n}%
\end{align}
To fix ideas, recall that for $Q_{100}$ it is,
\begin{align}
Q_{100}  &  =\frac{\partial}{\partial x^{1}}\left(  1-\frac{2m}{\sqrt
{(x^{1})^{2}+(x^{2})^{2}+(x^{3})^{2}}}\right) \nonumber\\
&  =-2m\frac{\partial}{\partial x^{1}}\left(  \frac{1}{\sqrt{(x^{1}%
)^{2}+(x^{2})^{2}+(x^{3})^{2}}}\right) \nonumber\\
&  =\frac{2mx^{1}}{[(x^{1})^{2}+(x^{2})^{2}+(x^{3})^{2}]^{\frac{3}{2}}}%
=\frac{2mx^{1}}{r^{3}}, \label{11n}%
\end{align}
which is \textit{non null} for $x^{1}\neq0$. \ Note that also that
$Q_{010}=Q_{001}=0$.

\subsection{$(M,\mathbf{\eta},\mathring{D})$, $(M,\mathbf{\eta},D)$, $(M,%
\slg
,D)$ and $(M,%
\slg
,\mathring{D})$}

From what has been said it is obvious that since $(M,\mathbf{\eta})$ and $(M,%
\slg
)$ are both orientable and time orientable, then $(M,\mathbf{\eta}%
,\mathring{D})$, $(M,%
\slg
,D)$ are part of the structures representing \ respectively Minkowski
spacetime and Schwarzschild spacetime. More precisely, $(M,%
\slg
,D,\tau_{%
\sslg
},\uparrow)$ represents in GR the gravitational field of a point mass with
world line given by $(t,0,0,0)$. As usual in GR this world line is left out of
the effective manifold \footnote{The manifold where Schwarzschild solution is
obtained is one with \textit{boundary}, i.e., it is $\mathbb{R}\times
\lbrack0,\infty)\times S^{2}$. The reason for that is that almost all
mathematical physicists use manifolds with \textit{boundary} in order to avoid
the use of distributions (generalized functions). Indeed, for a rigorous point
of view, taking into account that Einstein's equations are non linear we
cannot solve it using Schwartz distributions. To solve problems involving
singular distributions in GR in a rigorous way it is necessary to use
Colombeau theory of generalized functions as described, e.g., in
\cite{grosser}.}.

We claim that $(M,\mathring{D},%
\slg
)$ or $(M,\mathbf{\eta},D)$ are legitimate equivalent representations for the
gravitational field described in GR by the substructure $(M,%
\slg
,D)$. To find, e.g., the relation between the models $(M,%
\slg
,\mathring{D})$ and $(M,%
\slg
,D)$ it is necessary to recall that if in the bases $\{\partial_{\mu}\}$ for
$TM$ and $\{dx^{\nu}\}$ for $T^{\ast}M$, we have
\begin{align}
D_{\mbox{\tiny\boldmath$\partial$}_{\mu}}{\mbox{\boldmath$\partial$}}_{\nu}
&  =\Gamma_{\mu\nu}^{\rho}{\mbox{\boldmath$\partial$}}_{\rho},\nonumber\\
D_{\mbox{\tiny\boldmath$\partial$}_{\mu}}dx^{\alpha}  &  =-\Gamma_{\mu\nu
}^{\alpha}dx^{\nu}, \label{12n}%
\end{align}
and the Christoffel symbols are \textit{not} all null.\ Moreover, in the
spherical coordinates introduced above
\begin{align}
\mathring{D}_{\mbox{\tiny\boldmath$\partial$}_{\mu}^{\prime}}%
{\mbox{\boldmath$\partial$}}_{\nu}^{\prime}  &  =L_{\mu\nu}^{\prime\rho
}{\mbox{\boldmath$\partial$}}_{\rho}^{\prime},\text{ }\mathring{D}%
_{\partial_{\mu}^{\prime}}dx^{\prime\alpha}=-L_{\mu\nu}^{\prime\alpha
}dx^{\prime\nu}\nonumber\\
D_{\mbox{\tiny\boldmath$\partial$}_{\mu}^{\prime}}{\mbox{\boldmath$\partial$}}%
_{\nu}^{\prime}  &  =\Gamma_{\mu\nu}^{\prime\rho}\partial_{\rho}^{\prime
},D_{\partial_{\mu}^{\prime}}dx^{\prime\alpha}=-\Gamma_{\mu\nu}^{\prime\alpha
}dx^{\prime\nu} \label{13n}%
\end{align}
and the $L_{\mu\nu}^{\prime\rho}$ and $\Gamma_{\mu\nu}^{\prime\rho}$ are
\textit{not} all null. Now, $L_{\mu\nu}^{\prime\rho}$ and $\Gamma_{\mu\nu
}^{\prime\rho}$ are related by\footnote{See, e.g., Section 4.5.8 of
\cite{rodcap2007}.}:%

\begin{equation}
L_{\alpha\beta}^{\prime\rho}=\Gamma_{\alpha\beta}^{\prime\rho}+\frac{1}%
{2}S_{\alpha\beta}^{\prime\rho}, \label{14n}%
\end{equation}
where $S_{\alpha\beta}^{\prime\rho}$ are the components of the so called
\textit{strain tensor of the connection }$\mathring{D}$ relative to the
connection \ $D$. For the present case it is
\begin{equation}
S_{\alpha\beta}^{\prime\rho}=g^{\prime\rho\sigma}(Q_{\alpha\beta\sigma
}^{\prime}+Q_{\beta\sigma\alpha}^{\prime}-Q_{\sigma\alpha\beta}^{\prime}).
\label{15n}%
\end{equation}

Now, since in the Cartesian coordinates $L_{\alpha\beta}^{\rho}=0$, but not
all $\Gamma_{\alpha\beta}^{\rho}$ are null we get
\begin{equation}
\Gamma_{\alpha\beta}^{\rho}=-\frac{1}{2}S_{\alpha\beta}^{\rho} \label{16n}%
\end{equation}
and thus, e.g.,
\begin{equation}
g_{1\rho}\Gamma_{00}^{\rho}=-\frac{1}{2}S_{100}=\frac{1}{2}Q_{100}%
=\frac{mx^{1}}{r^{3}}. \label{17n}%
\end{equation}

\subsection{$\mathbf{A}^{\eta}$ as the Gravitational Field}

Note that using coordinates (Riemann normal coordinates $\{\xi^{\mu}\}$
covering $V\subset U\subset M$) naturally adapted to a reference frame
$Z\mathbf{\in}\sec TV$\footnote{For the mathematical definitions of reference
frames, naturally adapted coordinates to a reference frame and observers, see,
e.g., Chapter 5 of \cite{rodcap2007}.} in free fall according to GR
($D_{Z}Z=0,$ $d\alpha\wedge\alpha=0$, $\alpha=g(Z,\;)$) it is possible to put
the connection coefficients of the Levi-Civita connection $D$ of $%
\slg
$ equal to zero in all points of the world line of a free fall observer (an
\textit{observer} is here \textit{modelled} as an integral line $\sigma$ of a
reference frame $Z$, where $Z$ is a time like vector field pointing to the
future such that $\ \left.  Z\right\vert _{\sigma}=\sigma_{\ast}$).

In the Riemann normal coordinates system covering $U\subset M$, it is obvious
that not all the connection coefficients of the connection $\mathring{D}$
(that relative to $%
\slg
$ is a non metrical one) are null. Moreover, the nonmetricity tensor
$\mathbf{A}^{\eta}$ is not null and it represents in our model the true
gravitational field. Indeed, an observer following $\sigma$ does not fell any
force along its world line because the gravitational force represented by the
nonmetricity field $\mathbf{A}^{\eta}$ is compensated by an inertial force
represented by the non null connection coefficients\footnote{The explicit form
of the coefficients $L_{\mu\nu}^{\prime\prime\rho}$ may be found in Chapter 5
of \cite{rodcap2007}.} $L_{\mu\nu}^{\prime\prime\rho}$ of $\mathring{D}$ in
the basis $\{\frac{\partial}{\partial\xi^{\nu}}\}$.

The situation is somewhat analogous to what happens in any non inertial
reference frame which, of course, may be conveniently used in \textit{any}
Special Relativity problem (as e.g., in a rotating disc \cite{rodsha}), where
the connection coefficients of the Levi-Civita connection of $\mathbf{\eta}$
are not all null.

\begin{remark}
The theoretical definition of standard clocks of GR are reasonably well
realized by atomic clocks, i.e., under certain limits atomic clocks behave as
theoretically predicted \emph{(}see however \emph{\cite{rodoliv})} Note
however that atomic clocks are \textit{not} the standard clocks of the model
proposed here. \ We would say that the gravitational field \textit{distorts}
the period of the atomic clocks relative to the standard clocks of the
proposed model where gravity is represented by a nonmetricity
tensor\footnote{Schwinger \cite{schwinger} showed with very simple arguments
how the gravitational field \textit{distorts} the period of atomic clocks
making then to register the proper time predicted by GR. His arguments can be
easily adapted for the alternative models studied in this paper, because once
$%
\slg
$ is known experimentally we can determine $\eta$ with the\ mathematical
techniques described in \cite{rodcap2007}.}. But, who are the devices that now
materialize those concepts? Well, they may are \textit{paper concepts}, like
the notion of time in some Newtonian theories. They are defined and calculated
in order to make \textit{correct} predictions. However, given the status of
present technology we can easily imagine how to build devices for directly
realizing the standard clocks \emph{(}and rulers\emph{)} of the proposed model.
\end{remark}

\section{Conclusions}

In this paper we recalled two important results. The first is that a
gravitational field generated by a given energy-momentum distribution can be
represented by distinct geometrical structures (Lorentzian, teleparallel and
non null nonmetricity spacetimes). The second important result is that we can
even dispense all those geometrical structures and simply represent the
gravitational field as a field in the Faraday's sense living in Minkowski
spacetime. The explicit Lagrangian density for this theory has been discussed
and the field equations have been shown to be equivalent to Einstein's
equations. We hope that our study clarifies the real difference between
mathematical models and physical reality and leads people to think about the
real physical nature of the gravitational field (and also of the
electromagnetic field\footnote{As suggested, e.g., by the works of Laughlin
\cite{laughlin} and Volikov \cite{volovik}. Of course,, it may be necessary to
explore also other ideas, like e.g., existence of branes in string theory. But
this is a subject for another publication.})

As a final remark, we want to leave clear that after studying Einstein's
papers (and also papers by many others authors) on the use Riemann-Cartan
\footnote{The teleparallel spaces are particular cases of the Riemann-Cartan
ones. More on the classification of spacetime geometries may be found in
\cite{rodcap2007}.} to describe a classical unified theory of gravitation and
electromagnetism we became convinced that it seems impossible to represent the
electromagnetic field using a contraction of the torsion tensor (or the
torsion tensor) without introducing ad hoc hypothesis. Having said that we
recall that from time to time some authors return to the embryo of Einstein's
original idea claiming to have obtained an unified theory of gravitation and
electromagnetism using that tool. Among those theories that appeared in the
last few years some are completely worthless, since based in a very bad use of
Mathematical concepts, but some looks at least at a first sight interesting
enough (at least from the mathematical point of view) to deserve some
comments, which will be discussed elsewhere\footnote{We have in mind here: (a)
some papers by \ Vargas and Vargas and Torr, \ \cite{v,vt1,vt2,vt3} where they
claim that using the torsion tensor of some special Finsler connections it is
possible to obtain a unified theory of gravitation and electromagnetism (for
related papers on the subject \ by those authors, please consult
http://cartan-einstein-unification.com/published-papers.html); (b) a paper by
Unzicker where he claims to have found a description of electromagnetism
including the existence of quantized charges using teleparallel spacetimes
with defects \cite{unzicker0,unzicker1}.}.

\appendix

\section{The Levi-Civita and the Nunes Connection on $\mathring{S}^{2}$}

Consider $S^{2}$, an sphere of radius $\mathfrak{R}=1$ embedded in
$\mathbb{R}^{3}$. Let $(x^{1},x^{2})=(\vartheta,\varphi)$ $0<\vartheta<\pi$,
$0<\varphi<2\pi$, be the standard spherical coordinates of $S^{2}$, which
covers all the open set $U$ which is $S^{2}$ with the \textit{exclusion} of a
semi-circle uniting the \textit{north} and south \textit{poles}.

Introduce the \textit{coordinate bases}\textbf{ }
\begin{equation}
\{{\mbox{\boldmath$\partial$}}_{\mu}\},\{\theta^{\mu}=dx^{\mu}\}
\end{equation}
for $TU$ and $T^{\ast}U$. Next introduce the \textit{orthonormal
bases}\textbf{ }$\{%
\sle
_{\mathbf{a}}\},\{\theta^{\mathbf{a}}\}$ for $TU$ and $T^{\ast}U$ with%
\begin{subequations}
\begin{align}%
\sle
_{\mathbf{1}}  &  ={\mbox{\boldmath$\partial$}}_{1}\text{, }%
\sle
_{\mathbf{2}}=\frac{1}{\sin x^{1}}{\mbox{\boldmath$\partial$}}_{2}%
,\label{ba}\\
\theta^{\mathbf{1}}  &  =dx^{1}\text{, }\theta^{\mathbf{2}}=\sin x^{1}dx^{2}.
\end{align}
Then,%
\end{subequations}
\begin{align}
\lbrack%
\sle
_{\mathbf{i}},%
\sle
_{\mathbf{j}}]  &  =c_{\mathbf{ij}}^{\mathbf{k}}%
\sle
_{\mathbf{k}},\\
c_{\mathbf{12}}^{\mathbf{2}}  &  =-c_{\mathbf{21}}^{\mathbf{2}}=-\cot
x^{\mathbf{1}}.\nonumber
\end{align}
Moreover the metric $%
\slg
$\texttt{ }$\in\sec T_{2}^{0}S^{2}$ inherited form the ambient Euclidean
metric is:
\begin{align}%
\slg
&  =dx^{1}\otimes dx^{1}+\sin^{2}x^{1}dx^{2}\otimes dx^{2}\nonumber\\
&  =\theta^{\mathbf{1}}\otimes\theta^{\mathbf{1}}+\theta^{\mathbf{2}}%
\otimes\theta^{\mathbf{2}}.
\end{align}

The Levi-Civita connection $D$ of $%
\slg
$ has the following \ non null connections coefficients $\Gamma_{\mu\nu}%
^{\rho}$ in the coordinate basis (just introduced):%
\begin{align}
D_{\mbox{\tiny\boldmath$\partial$}_{\mu}}{\mbox{\boldmath$\partial$}}_{\nu}
&  =\Gamma_{\mu\nu}^{\rho}{\mbox{\boldmath$\partial$}}_{\rho},\nonumber\\
\text{ }\Gamma_{21}^{2}  &  =\Gamma_{\theta\varphi}^{\varphi}=\Gamma_{12}%
^{2}=\Gamma_{\varphi\theta}^{\varphi}=\cot\vartheta\text{, }\Gamma_{22}%
^{1}=\Gamma_{\varphi\varphi}^{\vartheta}=-\cos\vartheta\sin\vartheta.
\label{10x}%
\end{align}
Also, in the basis $\{%
\sle
_{\mathbf{a}}\}$, $D_{%
\sle
_{\mathbf{i}}}%
\sle
_{\mathbf{j}}=\omega_{\mathbf{ij}}^{\mathbf{k}}%
\sle
_{\mathbf{k}}$ and the non null coefficients are:
\begin{equation}
\omega_{\mathbf{21}}^{\mathbf{2}}=\cot\vartheta\text{, }\omega_{\mathbf{22}%
}^{\mathbf{1}}=-\cot\vartheta. \label{crist}%
\end{equation}

The torsion \ and the (Riemann) curvature tensors of $D$ (recall
Eq.(\ref{to op}) and Eq.(\ref{curv op}) are%

\begin{equation}
\mathcal{T}(\theta^{\mathbf{k}},%
\sle
_{\mathbf{i}},%
\sle
_{\mathbf{j}})=\theta^{\mathbf{k}}(\mathbf{\tau}(%
\sle
_{\mathbf{i}},%
\sle
_{\mathbf{j}}))=\theta^{\mathbf{k}}\left(  D_{%
\sle
_{\mathbf{j}}}%
\sle
_{\mathbf{i}}-D_{%
\sle
_{\mathbf{i}}}%
\sle
_{\mathbf{j}}-[%
\sle
_{\mathbf{i}},%
\sle
_{\mathbf{j}}]\right)  , \label{tolevi}%
\end{equation}%
\begin{equation}
\mathbf{R(}%
\sle
_{\mathbf{k}},\theta^{\mathbf{a}},%
\sle
_{\mathbf{i}},%
\sle
_{\mathbf{j}}\mathbf{)}=\theta^{\mathbf{a}}\left(  \left[  D_{%
\sle
_{\mathbf{i}}}D_{%
\sle
_{\mathbf{j}}}-D_{%
\sle
_{\mathbf{j}}}D_{%
\sle
_{\mathbf{i}}}-D_{[%
\sle
_{\mathbf{i}},\mathtt{\ }%
\sle
_{\mathbf{j}}]}\right]
\sle
_{\mathbf{k}}\right)  , \label{RIEMLEVI}%
\end{equation}
which results in\ $\mathcal{T}=0$ and that the non null components of
$\mathbf{R}$ are $R_{\mathbf{1\;21}}^{\;\mathbf{1}}=-R_{\mathbf{1\;12}%
}^{\;\mathbf{1}}=R_{\mathbf{1\;12}}^{\;\mathbf{2}}=-R_{\mathbf{1\;12}%
}^{\;\mathbf{2}}=-1$.

\begin{figure}
[ptb]
\begin{center}
\includegraphics[
natheight=2.260600in,
natwidth=2.250200in,
height=2.3004in,
width=2.29in
]%
{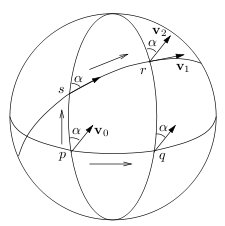}%
\caption{Levi-Civita and Nunes transport of a vector $\mathbf{v}_{0}$
satarting at $p$ through the paths $psr$ and $pqr$. Levi-Civita tranport
through psr leads to $\mathbf{v}_{1}$ whereas Nunes transport leads to
$\mathbf{v}_{2}$. Along $pqr$ both Levi-Civita and Nunes transport agree and
leads to $\mathbf{v}_{2}.$}%
\label{levitelle}%
\end{center}
\end{figure}

Since the Riemann curvature tensor is non null the parallel transport of a
given vector depends on the path to be followed. We say that a vector (say
$\mathbf{v}_{0}$) is parallel transported along a generic path
$\mathbb{R\supset}I\mapsto\gamma(s)\in\mathbb{R}^{3}$ (say, from $A=\gamma(0)$
to $B=\gamma(1)$) with tangent vector $\gamma_{\ast}(s)$ (at $\gamma(s)$) if
it determines a vector field $\mathbf{V}$ along $\gamma$ satisfying
\begin{equation}
D_{\gamma_{\ast}}\mathbf{V}=0,\label{lctransp}%
\end{equation}
and such that $\mathbf{V}(\gamma(0))=\mathbf{v}_{0}$. When the path is a
geodesic\footnote{We recall that a geodesic of $D$ also determines the minimal
distance (as given by the metric $%
\slg
$) between any two points on $S^{2}$.} of the connection $D$, i.e.,a curve
\ $\mathbb{R\supset}I\mapsto c(s)\in\mathbb{R}^{3}$ with tangent vector
$c_{\ast}(s)$ (at $c(s)$) satisfying%
\begin{equation}
D_{c_{\ast}}c_{\ast}=0,\label{geodlv}%
\end{equation}
the parallel transported vector along a $c$ forms a constant angle with
$c_{\ast}$. Indeed, from Eq.(\ref{lctransp}) it is $\gamma_{\ast}\underset{%
\sslg
}{\cdot}D_{\gamma_{\ast}}\mathbf{V}=0$. Then taking into account
Eq.(\ref{geodlv}) it follows that
\[
D_{\gamma_{\ast}}(\gamma_{\ast}\underset{%
\sslg
}{\cdot}\mathbf{V})=0.
\]
i.e., $\gamma_{\ast}\underset{%
\sslg
}{\cdot}\mathbf{V}=$ \emph{constant}$.$This is clearly illustrated in Figure
\ref{levitelle} (from \cite{bergmann}).%

\begin{figure}
[h]
\begin{center}
\includegraphics[
natheight=1.222800in,
natwidth=2.270100in,
height=2.9525in,
width=5.4163in
]%
{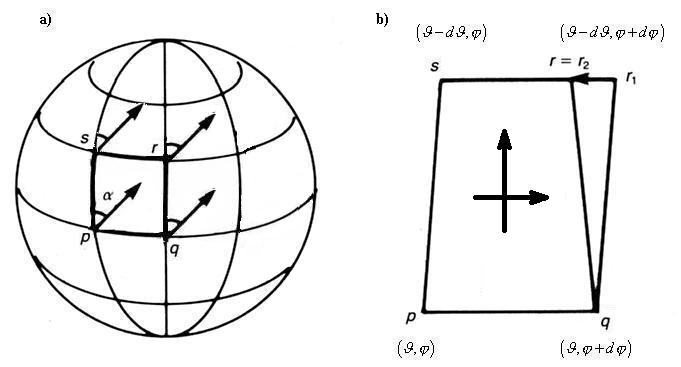}%
\caption{Characterization of the Nunes connection.}%
\label{figcolumbus}%
\end{center}
\end{figure}

Consider next the manifold $\mathring{S}^{2}$ $=\{S^{2}\backslash
$\textrm{north pole + south pole}$\}\subset\mathbb{R}^{3}$, which is our
sphere of radius $\mathfrak{R}=1$ but this time \textit{excluding} the north
and south poles. Let again \texttt{ }$%
\slg
\in\sec T_{2}^{0}\mathring{S}^{2}$ be the metric field on $\mathring{S}^{2}$
inherited from the ambient space $\mathbb{R}^{3}$ and introduce on
$\mathring{S}^{2}$ the Nunes (or navigator) connection\footnote{Pedro
Salacience Nunes (1502--1578) was one of the leading mathematicians and
cosmographers of Portugal during the Age of Discoveries. He is well known for
his studies in Cosmography, Spherical Geometry, Astronomic Navigation, and
Algebra, and particularly known for his discovery of loxodromic curves and the
nonius. Loxodromic curves, also called rhumb lines, are spirals that converge
to the poles. They are lines that maintain a fixed angle with the meridians.
In other words, loxodromic curves directly related to the construction of the
Nunes connection. A ship following a fixed compass direction travels along a
loxodromic, this being the reason why Nunes connection is also known as
navigator connection. Nunes discovered the loxodromic lines and advocated the
drawing of maps in which loxodromic spirals would appear as straight lines.
This led to the celebrated Mercator projection, constructed along these
recommendations. Nunes invented also the Nonius scales which allow a more
precise reading of the height of stars on a quadrant. The device was used and
perfected at the time by several people, including Tycho Brahe, Jacob Kurtz,
Christopher Clavius and further by Pierre Vernier who in 1630 constructed a
practical device for navigation. For some centuries, this device was called
nonius. During the 19$^{th}$ century, many countries, most notably France,
started to call it vernier. More details in
http://www.mlahanas.de/Stamps/Data/Mathematician/N.htm.} $\mathbf{\nabla}$
defined by the following parallel transport rule: a vector at an arbitrary
point of $\mathring{S}^{2}$ is parallel transported along a curve $\gamma$, if
it determines a vector field on $\gamma$ such that at any point of $\gamma$
the angle between the \ transported vector and the vector tangent to the
latitude line passing through that point is constant during the transport.
This is clearly illustrated in Figure \ref{figcolumbus}. and to distinguish
the Nunes transport from the Levi-Civita transport we ask also for the reader
to study with attention the caption of Figure (\ref{levitelle}).

\hspace{-8cm}%

We recall that\emph{ }from the calculation of the Riemann tensor $\mathbf{R}$
it follows that the structures $(\mathring{S}^{2},%
\slg
,D,\tau_{%
\sslg
})$ and also $(S^{2},%
\slg
,D,\tau_{%
\sslg
})$ are Riemann spaces of \textit{constant curvature}. We now show that the
structure $(\mathring{S}^{2},%
\slg
,\mathbf{\nabla},\tau_{%
\sslg
})$ is a \textit{teleparallel space\footnote{As recalled in Section 1, a
teleparallel manifold $M$ is characterized by the existence of global vector
fields which is a basis for $T_{x}M$ for any $x\in M$. The reason for
considering $\mathring{S}^{2}$ for introducing the Nunes connection is that as
well known (see, e.g., \cite{docarmo}) $S^{2}$ does not admit a continuous
vector field that is nonnull at on points of it. }}, with zero Riemamn
curvature tensor, but non zero torsion tensor.

Indeed, from Figure \ref{figcolumbus} it is clear that (a) if a vector is
transported along the \textit{infinitesimal} quadrilateral $pqrs$ composed of
latitudes and longitudes, first starting from $p$ along $pqr$ and then
starting from $p$ along $psr$ the parallel transported vectors that result in
both cases will coincide (study also the caption of Figure (\ref{levitelle}).

Now, the vector fields $%
\sle
_{\mathbf{1}}$ and $%
\sle
_{\mathbf{2}}$ in Eq.(\ref{ba}) define a basis for each point $p$ of
$T_{p}\mathring{S}^{2}$ and $\mathbf{\nabla}$ is clearly characterized by:
\begin{equation}
\mathbf{\nabla}_{%
\sle
_{\mathbf{j}}}%
\sle
_{\mathbf{i}}=0. \label{6x}%
\end{equation}

The \ components of curvature operator are:%

\begin{equation}
\mathbf{\bar{R}(}%
\sle
_{\mathbf{k}},\theta^{\mathbf{a}},%
\sle
_{\mathbf{i}},%
\sle
_{\mathbf{j}}\mathbf{)}=\theta^{\mathbf{a}}\left(  \left[  \mathbf{\nabla}_{%
\sle
_{\mathbf{i}}}\mathbf{\nabla}_{%
\sle
_{\mathbf{j}}}-\mathbf{\nabla}_{%
\sle
_{\mathbf{j}}}\mathbf{\nabla}_{%
\sle
_{\mathbf{i}}}-\mathbf{\nabla}_{[%
\sle
_{\mathbf{i}},%
\sle
_{\mathbf{j}}]}\right]
\sle
_{\mathbf{k}}\right)  =0,
\end{equation}
and the torsion operation ( recall Eq.(\ref{top})) $\mathbf{\bar{\tau}}$ is:%
\begin{align}
\mathbf{\bar{\tau}}(%
\sle
_{\mathbf{i}},%
\sle
_{\mathbf{j}})  &  =\mathbf{\nabla}_{%
\sle
_{\mathbf{j}}}%
\sle
_{\mathbf{i}}-\mathbf{\nabla}_{%
\sle
_{\mathbf{i}}}%
\sle
_{\mathbf{j}}-[%
\sle
_{\mathbf{i}},%
\sle
_{\mathbf{j}}]\nonumber\\
&  =[%
\sle
_{\mathbf{i}},%
\sle
_{\mathbf{j}}],
\end{align}
which gives for the components of the torsion tensor, $\bar{T}_{\mathbf{12}%
}^{\mathbf{2}}=-\bar{T}_{\mathbf{12}}^{\mathbf{2}}=\cot\vartheta$ . It follows
that $\mathring{S}^{2}$ considered as part of the structure $(\mathring{S}%
^{2},%
\slg
,\mathbf{\nabla},\tau_{%
\sslg
})$ is\textit{ flat }(but has torsion)!

If you still need more details to grasp this last result, consider Figure
\ref{figcolumbus}(b) which shows the standard parametrization of the points
$p,q,r,s$ in terms of the spherical coordinates introduced above. According to
the geometrical meaning of torsion, its value at a given point is determined
by calculating the difference between the (infinitesimal)\footnote{This
wording, of course, means that those vectors are identified as elements of the
appropriate tangent spaces.} vectors $pr_{1}$and $pr_{2}$. If the vector $pq$
is transported along $ps$ one get (recalling that $\mathfrak{R}=1\mathfrak{)}$
the vector $\mathbf{v}=sr_{1}$ such that $\left\vert
\slg
(\mathbf{v},\mathbf{v})\right\vert ^{\frac{1}{2}}=\sin\vartheta\triangle
\varphi$. On the other hand, if the vector $ps$ is transported along $pq$ one
get the vector $qr_{2}=qr$. Let $\mathbf{w}=sr$. Then,%

\begin{equation}
\left\vert
\slg
(\mathbf{w},\mathbf{w})\right\vert =\sin(\vartheta-\triangle\vartheta
)\triangle\varphi\simeq\sin\vartheta\triangle\varphi-\cos\vartheta
\triangle\vartheta\triangle\varphi,
\end{equation}
Also,%

\begin{equation}
\mathbf{u}=r_{1}r_{2}=-u(\frac{1}{\sin\vartheta}\frac{\partial}{\partial
\varphi})\text{, }u=\left\vert
\slg
(\mathbf{u},\mathbf{u})\right\vert =\cos\vartheta\triangle\vartheta
\triangle\varphi.
\end{equation}
Then, the connection $\mathbf{\nabla}$ of the structure $(\mathring{S}^{2},%
\slg
,\mathbf{\nabla},\tau_{%
\sslg
})$ has a non null torsion tensor $\mathcal{\bar{T}}$. \ Indeed, the component
of $\mathbf{u}=r_{1}r_{2}$ in the direction $\partial/\partial\varphi$ is
precisely $\bar{T}_{\vartheta\varphi}^{\varphi}\triangle\vartheta
\triangle\varphi$. So, one get (recalling that $\mathbf{\nabla}_{\partial
j}\partial_{i}=\Gamma_{ji}^{k}\partial_{k}$)
\begin{equation}
\bar{T}_{\vartheta\varphi}^{\varphi}=\left(  \Gamma_{\vartheta\varphi
}^{\varphi}-\Gamma_{\varphi\vartheta}^{\varphi}\right)  =-\cot\theta.
\end{equation}
To end this Appendix it is worth to show that $\mathbf{\nabla}$ is metrical
compatible, i.e., that $\mathbf{\nabla}%
\slg
=0$. Indeed, we have:%
\begin{align}
0  &  =\text{ }\mathbf{\nabla}_{%
\sle
_{\mathbf{c}}}%
\slg
(%
\sle
_{\mathbf{i}},%
\sle
_{\mathbf{j}})=(\mathbf{\nabla}_{%
\sle
_{\mathbf{c}}}%
\slg
)(%
\sle
_{\mathbf{i}},%
\sle
_{\mathbf{j}})+\mathtt{\ }%
\slg
(\mathbf{\nabla}_{%
\sle
_{\mathbf{c}}}%
\sle
_{\mathbf{i}},%
\sle
_{\mathbf{j}})+\mathtt{\ }%
\slg
(%
\sle
_{\mathbf{i}},\mathbf{\nabla}_{%
\sle
_{\mathbf{c}}}%
\sle
_{\mathbf{j}})\nonumber\\
&  =(\mathbf{\nabla}_{%
\sle
_{\mathbf{c}}}%
\slg
)(%
\sle
_{\mathbf{i}},%
\sle
_{\mathbf{j}}).
\end{align}

\begin{remark}
This Appendix shows clearly that we cannot \textit{mislead} the Riemann
curvature tensor of a connection with the fact that the manifold where that
connection is defined may be bend\emph{\footnote{Bending of surfaces embedded
in $\mathbb{R}^{3}$ is adequately characterized by the so called shape
operator \ discussed, e.g., in\cite{oneill}. \ For the case of hypersurfaces
(vector manifolds) embedded in $\mathbb{R}^{n}$ see \cite{heso}.}} as a
surface in an Euclidean manifold where it is embedded. Neglecting this fact
may generate a lot of \textit{wishful} thinking when one comes to the
interpretation of curvature \emph{(}and torsion\emph{) }in gravitational theories.
\end{remark}


\begin{thebibliography}{99}                                                                                               %


\bibitem {bergmann}Bergmann, P. G., \textit{The Riddle of Gravitation}, Dover,
New York, 1992.

\bibitem {cartan}Cartan, E., A Generalization of the Riemann Curvature and the
Spaces with Torsion, \textit{Comptes Rendus Acad. Sci. }\textbf{174}, 593-596 (1922).

\bibitem {choquet}Choquet-Bruhat, Y., DeWitt-Morette, C. and Dillard-Bleick,
M., \textit{Analysis, Manifolds and Physics} (revised edition), North Holland
Publ. Co., Amsterdam, 1982.

\bibitem {clarke}Clarke, C. J. S., On the Global Isometric Embedding of
Pseudo-Riemannian Manifolds, \textit{Proc. Roy. Soc. A} \textbf{314}, 417-428 (1970).\ 

\bibitem {coll}Coll, B., A Universal Law of Gravitational Deformation for
General Relativity, \textit{Proc. of the Spanish Relativistic Meeting, EREs},
Salamanca, Spain,1998.

\bibitem {deandrade}de Andrade, V. C., Arcos, H. I., and Pereira, J. G.,
Torsion as an Alternative to Curvature in the Description of Gravitation, PoS
WC, 028 922040, \texttt{[arXiv:gr-qc/0412034}]

\bibitem {debewer}Debewer, R., \textit{\'{E}lie Cartan- Albert Einstein:
Letters on Absolute Parallelism}, Princeton University Press, Princeton, 1979.

\bibitem {docarmo}do Carmo, M. P., \textit{Riemannian Geometry}%
,\ Birkhh\"{a}user, Boston 1992.

\bibitem {eddington}Eddington, A. S., \textit{The Mathematical Theory of
Relativity }(3rd edn)., Chelsea, New York, 1975.

\bibitem {einstein1}Einstein, A., Unified Field Theory of Gravitation and
Electricity, \textit{Session Report of Prussian Acad. Sci.}, 414-419, July 25 (1925).

\bibitem {einstein2}Einstein, A., Unified Field Theory of Gravitation and
Electricity, \textit{Session Report of the Prussian Acad. Sci.}, 217-221, June
7th (1928).

\bibitem {einstein3}Einstein, A., New Possibility for a Unified Field Theory
of Gravitation and Electricity, \textit{Session Report of the Prussian Acad.
Sci}., 224-227, June 14th (1928)

\bibitem {einstein4}Einstein, A., Unified Field Theory Based on Riemannian
Metrics and Distant Parallelism, \textit{Math. Ann.} \textbf{102}, 685-697 (1930).

\bibitem {fr}Fern\'{a}ndez, V. V., and Rodrigues, W. A. Jr.,
\textit{Gravitation as a Plastic Deformation of the Lorentz Vacuum,} in
preparation, (2009)

\bibitem {Feynman}Feynman, R. P., Morinigo, F. B. and Wagner, W. G., (edited
by Hatfield, B.), \textit{Feynman Lectures on Gravitation}, Addison-Wesley
Publ. Co., Reading, MA, 1995.

\bibitem {frankel}Frankel, T., The Geometry of Physics (second edition),
Cambridge Univ. Press, Cambridge, 1997.

\bibitem {geroch}Geroch, R. Spinor Structure of Space-Times in General
Relativity I, \textit{J. Math. Phys.} \textbf{9}, 1739-1744 (1988).

\bibitem {goenner}Goenner, H. F. M, On the History of Unified Field Theories,
\textit{Living Rev. in Relativity} \textbf{7},
lrr-2004-2(2004),{\small [http://relativity.livingreviews.org/Articles/lrr-2004-2]}%


\bibitem {gris}Grishchuk, L.~P, Some Uncomfortable Thoughts on the Nature of
Gravity, Cosmology, and the Early Universe (to appear in \textit{Space
Sciences Reviews)}, \texttt{[arXiv:0903.4395] }

\bibitem {grosser}Grosser, M. , Kunzinger, M. , Oberguggenberger, M. and
Steinbawer, R., \textit{Geometric Theory of Generalized Functions with
Applications to Relativity}, Mathematics and its Applications vol.
\textbf{537},Kluwer, Dordrecht, 2001.

\bibitem {hermann}Hermann, R., \textit{Ricci and Levi-Civita's Tensor Analysis
Paper. Translation, Comments and Additional Material. }Lie Groups: History,
Frontiers\& Applications vol.\textit{ }\textbf{II}, Math. Sci. Press,
Brookline, MA, 1975.

\bibitem {heso}Hestenes, D. and Sobczyck, G.,\textit{ Clifford Algebra to
Geometrical Calculus}, D. Reidel Publ. Co., Dordrecht, 1984.

\bibitem {landau}Landau, L.D. and Lifshitz, E. M., \textit{The Classical
Theory of Fields} (fourth revised english edition), Pergamon Presss, New York, 1975.

\bibitem {laughlin}Laughlin, R.B., A Different Universe: Reinventing Physics
from the Bottom, Basic Books, New York, 2005.

\bibitem {logunov1}Logunov, A. A., Mestvirishvili, \textit{The Relativistic
Theory of Gravitation}, Mir Publ., Moscow, 1989.

\bibitem {notterod}Notte-Cuello, E. A. \ and Rodrigues, W. A. Jr., A Maxwell
Like Formulation of Gravitational Theory in Minkowski Spacetime, \textit{Int.
J. Mod. Phys. D} \textbf{16}, 1027-1041 (2007).

\bibitem {notterod1}Notte-Cuello, E. A. \ and Rodrigues, W. A. Jr., Freud's
Identity of Differential Geometry, the Einstein-Hilbert Equations and the
Vexatious Problem of the Energy-Momentum Conservation in GR, \textit{Adv.
Appl. Cliford Algebras }\textbf{19, }113-145 (2009).

\bibitem {ohanian}Ohanian, CH. C. and Ruffini, R., \textit{Gravitation and
Spacetime} (second edition), W. W. Norton \& Co., New York, 1994.

\bibitem {oneill}O'Neill, B., \textit{Elementary Differential Geometry},
Academic Press, New York, 1966.

\bibitem {poincare}Poincar\'{e}, H., \textit{La Science et L' Hypoth\`{e}se},
Flamarion, Paris, 1902.

\bibitem {ricci1}Ricci-Curbastro, G., \textquotedblleft Sulla Teoria degli
Iperspazi\textquotedblright\ Rend. Acc. Lincei Serie IV, 232--237 (1895).
Reprinted in : Gregorio Ricci-Curbastro, Opere, vol. \textbf{I}, 431--437,
Edizioni Cremonense, Roma,1956.

\bibitem {ricci2}Ricci-Curbastro, G., Dei sistemi di congruenze ortogonali in
una variet\`{a} Qualunque, \textit{Mem. Acc. Lincei }Serie 5 (vol.
\textbf{II}) 276--322, (1896). Reprinted in: Gregorio Ricci-Curbastro, Opere
Vol II, 1--61, Editore Cremonense, Roma, 1957.

\bibitem {riccicivita}Ricci, G., and Levi-Civita, T., M\'{e}thodes de Calcul
Diff\'{e}rentielle Absolu et leurs Applications, \textit{Mathematische
Annalen} \textbf{54}, 125--201 (1901).

\bibitem {rolrod}da Rocha, R., and Rodrigues, W. A. Jr., Gauge Fixing in the
Maxwell Like Gravitational Theory in Minkowski Spacetime and in the Equivalent
Lorentzian Spacetime ,\texttt{[arXiv:0806.4129]}

\bibitem {rodoliv}Rodrigues, W.A. Jr. and Capelas de Oliveira, E., A Comment
on the Twin Paradox and the Hafele Keating Experiment , \textit{Phys. Letters
A} \textbf{140}, 479 484 (1989).

\bibitem {rodcap2007}Rodrigues, W.A. Jr.,and Capelas de Oliveira, E.,
T\textit{he Many Faces of Maxwell, Dirac and Einstein Equations. A Clifford
Bundle Approach. }Lecture Notes in Physics\textit{ }\textbf{722}, Springer,
Heidelberg, 2007.

\bibitem {rodsha}Rodrigues, W. A. Jr., and Sharif, M., Rotating Frames in RT:
Sagnac's Effect in SRT and other Related Issues, \textit{Found. Phys.}
\textbf{31}, 1767-1784 (2001).

\bibitem {rodquin}Rodrigues, W. A. Jr., and Souza ,Q. A. G., An Ambiguous
Statement Called 'Tetrad Postulate' and the Correct Field Equations Satisfied
by the Tetrad Fields, \textit{Int.J.Mod.Phys D} \textbf{14}, 2095-2150
(2005),\ \texttt{[arXiv:math-ph/0411085]}

\bibitem {sawu}Sachs, R. K., and Wu, H., \textit{General Relativity for
Mathematicians}, Springer-Verlag, New York 1977.

\bibitem {schu}Sch\"{u}cking, E. L., \textit{Einstein's Apple and Relativity's
Gravitational Field}, \texttt{[arXiv:0903.3768]}

\bibitem {schwinger}Schwinger, J., Particles, \textit{Sources and Fields, vol.
1}, Addison-Wesley Publ. Co., Reading, MA, 1970.

\bibitem {sab}Szabados, L. B., Quasi-Local Energy-Momentum and Angular
Momentum in GR: A Review Article, \textit{Living Reviews in Relativity}, [http://www.livingreviews.org/lrr-2004-4]

\bibitem {thirring}Thirring, W., An Alternative Approach to the Theory of
Gravitation, \textit{Ann. Phys.} \textbf{16}, 96-117 (1961).

\bibitem {unzicker}Unzicker, A., and Case, T.,Translation of Einstein's
Attempt of a Unified Field Theory with Teleparallelism,
\texttt{[arXiv:physics/0503046]}

\bibitem {unzicker0}Unzicker, A., What Can Physics Learn from Continuum
Mechanics, \texttt{[arXiv:gr-qc/0011064v1]}

\bibitem {unzicker1}Unzicker, A., Teleparallel Space-Time with Defects yields
Geometrization of Electrodynamics with Quantized Charges,
\texttt{[arXiv:gr-qc/9612061v2] }

\bibitem {v}Vargas, J. G., Geometrization of the Physics with Teleparallelism
(I): The Classical Interactions, \textit{Found. Phys.} \textbf{22}, 507-526 (1992).

\bibitem {vt1}Vargas, J. G., Torr, D. G., and Lecompte, A., Geometrization of
the Physics with Teleparallelism (II): Towards a Fully Geometric Dirac
Equation, \textit{Found. Phys.} \textbf{22}, 527-547 (1992).

\bibitem {vt2}Vargas, J. G., and Torr, D. G., Finslerian Structures: The
Cartan-Clifton Method of the Moving Frame, \textit{J. Math. Phys.}
\textbf{34}, 4898-4913 (1993).

\bibitem {vt3}Vargas, J. G., and Torr, D. G., The Cornerstone Role of the
Torsion in Finslerian Physical Theories, \textit{Gen. Rel. Grav.} \textbf{27},
629-644 (l995).

\bibitem {vt4}Vargas, J. G., andTorr, D. G., Is Electromagnetic Gravity
Control Possible?, in El-Genk, M. S. (editor),\ Proceedings of the 2004 Space
Technology and Applications International Forum (STAIF 2004), \textit{AIP
Conference Proceedings}, 1206-1213 ( 2004).

\bibitem {volovik}Volovik, G. E., \textit{The Universe in a Helium Droplet},
Clarendon Press, Oxford (2003),

\bibitem {weeks}Weeks, J. R., \textit{The Shape of Space }(second edition),
Marcel Decker Inc., New York, 2002.

\bibitem {weinberg}Weinberg, S., \textit{Gravitation and Cosmology}, J. Wiley
and Sons, Inc., New York, 1972.

\bibitem {wein}Weitzenb\"{o}ck, R., Differentialinvarianten in der
Einsteinschen Theorie des Fernparallelismus, \textit{Sitzungsber. Preuss.
Akad. Wiss.}\textbf{ 26}, 466-474, (1928).

\bibitem {Zorawski}Zorawski, M., \textit{Theorie Mathematiques des
Dislocations}, Dunod, Paris, 1967.
\end{thebibliography}
\end{document}